\documentclass[preprint,aps, amsmath,amssymb]{revtex4-1}

\usepackage{MnSymbol}
\usepackage[utf8]{inputenc}
\usepackage{soulutf8}
\usepackage{graphicx}
\usepackage{epstopdf}
\usepackage{amsmath}
\usepackage{slashed}
\usepackage{xcolor}
\usepackage{mathtools}
\usepackage{slashed}

\usepackage{array,multirow}
\usepackage{booktabs}
\usepackage{adjustbox} 
\usepackage{rotating}
\usepackage{blindtext}
\usepackage{physics}

\begin{document}

\title{Hadron spectroscopy using the light-front holographic Schr\"odinger equation and the 't Hooft equation}

\author{Mohammad Ahmady}
\email{mahmady@mta.ca}
\affiliation{\small Department of Physics, Mount Allison University, Sackville, New Brunswick, Canada, E4L 1E6}

\author{Satvir Kaur}
\email{satvirkaur578@gmail.com}
\affiliation{\small Department of Physics, Dr. B.R. Ambedkar National Institute of Technology, Jalandhar 144011, India}

\author{Sugee Lee MacKay}
\email{slmackay@mta.ca}
\affiliation{\small Department of Physics, Mount Allison University, Sackville, New Brunswick, Canada, E4L 1E6}

\author{Chandan Mondal}
\email{mondal@impcas.ac.cn}
\affiliation{\small Institute of Modern Physics, Chinese Academy of Sciences, Lanzhou 730000, China $\&$ School of Nuclear Science and Technology, University of Chinese Academy of Sciences, Beijing 100049, China}

\author{Ruben Sandapen}
\email{ruben.sandapen@acadiau.ca}
\affiliation{\small Department of Physics, Acadia University, Wolfville, Nova-Scotia, Canada, B4P 2R6}

\begin{abstract} 
Light-front holographic QCD  provides a successful first approximation to hadron spectroscopy in the chiral limit of $(3+1)$-dim light-front QCD, where a holographic Schr\"odinger-like equation, with an emerging confining scale, $\kappa$, governs confinement in the transverse direction.  In its supersymmetric formulation, light-front holography predicts that each baryon has two superpartners: a meson and a tetraquark, with  their degenerate masses being generated by the same scale, $\kappa$. In nature, this mass degeneracy is lifted by chiral symmetry breaking and longitudinal confinement. In this paper, we show that the latter can be successfully captured by the 't Hooft equation of $(1+1)$-dim, large $N_c$, QCD.  Together, the holographic Schr\"odinger equation and the 't Hooft equation, provide a good global description of the data across the full hadron spectrum with a universal $\kappa$.  
\end{abstract}

\maketitle

\section{Introduction}
In light-front $(3+1)$-dim QCD, the mass of a quark-antiquark meson is given by \cite{Brodsky:2014yha}
\begin{equation}
	M^2=\int \mathrm{d} x \mathrm{d}^2 \mathbf{b}_{\perp} \Psi^*(x,\mathbf{b}_\perp) \left[ -\frac{\nabla_{b_\perp}^2}{x(1-x)} +\frac{m^2_q}{x} + \frac{m^2_{\bar{q}}}{1-x} \right] \Psi(x,\mathbf{b}_\perp) + \mathrm{interactions}\;,
\label{M}
\end{equation}
where $\Psi(x,\mathbf{b}_\perp)$ is the meson light-front wave function, $x$ the light-front momentum fraction carried by the quark, and $\mathbf{b}_\perp$ (or $b_\perp e^{i\varphi}$ in polar representation) the transverse distance between the quark and the antiquark. In Equation \eqref{M}, $m_q$ and $m_{\bar{q}}$ are the effective quark and antiquark masses respectively \cite{Brodsky:2014yha}. The ``interactions" contain all the complicated QCD dynamics of a bound state, including the effect of higher Fock sectors on the valence sector. 

Introducing the light-front variable,
\begin{equation}
	\boldsymbol{\zeta}=\sqrt{x(1-x)} \mathbf{b}_\perp \;\,
\label{zeta}
\end{equation}
allows a factorization of the wave function into a transverse mode, $\phi(\zeta)$, and a longitudinal mode, $X(x)$: 
\begin{equation}
	\Psi (x, \zeta, \varphi)= \frac{\phi (\zeta)}{\sqrt{2\pi \zeta}}	 e^{i L \varphi}
 X(x)\,
	\label{full-mesonwf}
\end{equation}
with $L=|L_z^{\mathrm{max}}|$ being the light-front orbital angular momentum, and $X(x)=\sqrt{x(1-x)} \chi(x)$. Equation \eqref{M} then becomes
\begin{equation}
	M^2=M_\perp^2 + M_\parallel^2\,
\end{equation}
where
\begin{equation}
	M_\perp^2=\int \mathrm{d}^2 \boldsymbol{\zeta} \phi^*(\zeta) \left[-\frac{d^2}{d \zeta^2}+\frac{4L^2-1}{4 \zeta^2}+U_\perp(\zeta) \right] \phi(\zeta)\;,
\label{Mperp}
\end{equation}
and
\begin{equation}
	M_\parallel^2=\int \mathrm{d} x \chi^*(x) \left[\frac{m_q^2}{x}+\frac{m_{\bar{q}}^2}{1-x} +U_\parallel(x) \right] \chi(x) \;,
\label{Mpara}
\end{equation}
with
\begin{equation}
	\int \mathrm{d} x |\chi(x)|^2 = \int \mathrm{d}^2 \boldsymbol{\zeta} |\phi(\zeta)|^2=1 \;.
\end{equation}
Here the QCD bound-state dynamics are encoded in the potentials $U_\perp(\zeta)$ and $U_\parallel(x)$. Their exact derivation from first principles remains an open question.

Light-front holography \cite{Brodsky:2006uqa,deTeramond:2005su,deTeramond:2008ht,Brodsky:2014yha} considers the chiral limit, i.e., massless quarks, and neglects longitudinal confinement, thus implying that $M_\parallel=0$. Remarkably, the form of the transverse confinement potential, $U_\perp (\zeta)$, is uniquely fixed by the underlying conformal symmetry and a holographic mapping to anti-deSitter $\mathrm{AdS}_5$, resulting in \cite{Brodsky:2006uqa,deTeramond:2005su,deTeramond:2008ht,Brodsky:2014yha}
\begin{equation}
	U_\perp^{\mathrm{LFH}}(\zeta)=\kappa^4 \zeta^2 + 2\kappa^2(J-1) \;
\label{U-LFH}
\end{equation}
where $J=L+S$. In this holographic mapping, the variable $\zeta$ maps onto the fifth dimension of $\mathrm{AdS}_5$.  Equation \eqref{Mperp}, which can be rewritten as
\begin{equation}
	 \left(-\frac{d^2}{d \zeta^2}+\frac{4L^2-1}{4 \zeta^2}+U^{\mathrm{LFH}}_\perp(\zeta)\right)\phi(\zeta)= M_\perp^2 \phi(\zeta) \;,
\end{equation}
maps onto the wave equation for the amplitude of spin-$J$ modes propagating in $\mathrm{AdS}_5$ modified by a quadratic dilaton field. In light-front holography, the longitudinal mode, $X(x)$, is hence not dynamical, i.e., undetermined by Equation \eqref{Mpara}. It is instead fixed by the holographic mapping of the electromagnetic (or gravitational) pion form factor in physical spacetime and $\mathrm{AdS}_5$ \cite{Brodsky:2007hb,Brodsky:2008pf}, resulting in $\chi(x)=1$, i.e.,
\begin{equation}
	X(x)=\sqrt{x(1-x)} \;.
\label{X-LFH}
\end{equation}

The supersymmetric formulation of light-front holography \cite{Dosch:2015bca,Dosch:2016zdv,Nielsen:2018uyn,Nielsen:2018ytt,Brodsky:2016rvj} provides a unified description of baryons and mesons/tetraquarks where each baryon (viewed as a quark-diquark system) possesses two superpartners: a (quark-antiquark) meson and a (diquark-antidiquark) tetraquark. This supersymmetric connection stems from the fact that a diquark can be in the same $\textrm{SU}_c(3)$ representation as an antiquark, and an antidiquark can be in the same $\textrm{SU}_c(3)$ representation as a quark. The supersymmetric holographic Schr\"odinger equation reads \cite{Nielsen:2018uyn}	
\begin{equation}
H \ket{\phi}=M_\perp^{2} \ket{\phi}
\label{SuSy-Holographic-SE}
\end{equation}
where 
 \begin{equation}
	H=
	\begin{pmatrix}
		-\frac{d^2}{d \zeta^2}+\frac{4L_M^2-1}{4 \zeta^2}+U_{M}(\zeta) & 0\\
		0 & -\frac{d^2}{d \zeta^2}+\frac{4L_B^2-1}{4 \zeta^2}+U_{B}(\zeta)
	\end{pmatrix}
\end{equation}
with
\begin{equation}
	U_M(\zeta)=\kappa^4 \zeta^2 + 2\kappa^2 (L_M+S_M -1)\,,
\label{UM}
\end{equation}
\begin{equation}
	U_B(\zeta)=\kappa^4 \zeta^2 + 2\kappa^2(L_B+S_D)
	\label{UB}
\end{equation}
where $S_M$ is the total quark-antiquark spin and $S_D$ is the diquark spin. The $4$-plet, $\ket{\phi}$, is given by
\begin{equation}
	\ket{\phi}=
	\begin{pmatrix}
		\phi_{M}(L_M=L_B+1) & \psi^{-}(L_B+1)\\
		\psi^{+}(L_B) & \phi_{T}(L_T=L_B)
	\end{pmatrix}
\end{equation}
where $\psi^{+}$ and $\psi^{-}$ are the two components of the baryon wave function and $\phi_{M/T}$ is the meson/tetraquark wave function. Equation \eqref{SuSy-Holographic-SE} admits analytical solutions, with its eigenvalues given by
\begin{equation}
	M_{\perp,M}^2=4\kappa^2\left(n_\perp + L_M + \frac{S_M}{2}\right) \;,
	\label{MTM}
\end{equation}
\begin{equation}
	M_{\perp,B}^2=4\kappa^2\left(n_\perp + L_B  + \frac{S_D}{2} + 1\right) \;,
	\label{MTB}
\end{equation}
and
\begin{equation}
	M_{\perp,T}^2=4\kappa^2\left(n_\perp + L_T + \frac{S_T}{2}+1\right) \,,
	\label{MTT}
\end{equation}
where $n_\perp$ is the principal quantum number that emerges when solving the holographic Schr\"odinger equation, and $S_T$ is total diquark-antidiquark spin. 

 It follows that baryons with quantum numbers, $L_B=L_M-1$ and $S_D=S_M$, are superpartners to mesons with quantum numbers, $L_M$ and $S_M$, and tetraquarks with quantum numbers $S_T=S_D$ and $L_T=L_B$. Hence, within a given family of superpartners, we can write $S\equiv S_D=S_M=S_T$. Note that the lowest-lying mesons with $n_\perp=L_M=0$ do not have a baryonic superpartner. Furthermore, pseudoscalar mesons with $n_\perp=L_M=S_M=0$, like the pion and the kaon, are predicted to be massless, just as expected in the chiral limit of QCD.

\section{Chiral symmetry breaking and longitudinal confinement}
The effects of nonzero quark masses were originally taken into account using a prescription by Brodsky and de T\'eramond (BdT) \cite{Brodsky:2008pg}, which relies on the fact that the holographic ground state wave function depends on the invariant mass of the $q\bar{q}$ pair. Indeed, in momentum space,
\begin{equation}
	\tilde{\phi}(x,\mathbf{k}) \propto \frac{1}{\sqrt{x(1-x)}} \exp \left(-\frac{M_{q\bar{q}}^2}{2\kappa^2} \right)
\end{equation}
where 
\begin{equation}
	M_{q\bar{q}}^2=\frac{\mathbf{k}^2}{x(1-x)}
\end{equation}
is the invariant mass of the $q\bar{q}$ pair, with $\mathbf{k}$ being the transverse momentum of the quark. For massive quarks, the invariant mass becomes
\begin{equation}
	M_{q\bar{q}}^2=\frac{\mathbf{k}^2}{x(1-x)} +\frac{m^2_q}{x} + \frac{m^2_{\bar{q}}}{1-x}\;.
\end{equation}
Thus, after Fourier transforming back to position space, the BdT prescription amounts to a modification of the longitudinal mode:
\begin{equation}
X_{\mathrm{BdT}}(x)= \sqrt{x(1-x)} \chi_{\mathrm{BdT}}(x)
	\label{BT-X}
\end{equation}
where 
\begin{equation}
\chi_{\mathrm{BdT}}(x)=\exp\left(-\frac{(1-x)m_q^2 + xm^2_{\bar{q}}}{2\kappa^2x(1-x)} \right) \,.
\label{chi-BdT}
\end{equation}

Treating the kinetic energy of massive quarks as a first-order perturbation to the holographic potential, Equation \eqref{U-LFH}, the resulting first-order shift to $M=M_\perp$ due to quark masses is then given by
\begin{equation}
	\Delta M_{\mathrm{BdT}}^2= \int \mathrm{d} x \chi_{\mathrm{BdT}}^2(x) \left(\frac{m_q^2}{x}+\frac{m_{\bar{q}}^2}{1-x} \right)\,,
\label{BT-shift}
\end{equation}
so that the pion mass, $M_\pi=\Delta M_{\mathrm{BdT}}$. Similarly, the kaon mass, $M_K=\Delta M_{\mathrm{BdT}}$ when the strange quark is taken into account. Equation \eqref{BT-shift} can be generalized for baryons and tetraquarks \cite{Brodsky:2016rvj}:
\begin{equation}
	\Delta M_{\mathrm{BdT}}^2 = \frac{\lambda^2}{F}\frac{\mathrm{d}F[\lambda]}{\mathrm{d}\lambda}
\label{BdT-shift-general}
\end{equation}
where 
\begin{equation}
F[\lambda]=\int_0^1 \prod_{i=1}^n \mathrm{d} x_i e^{-\frac{1}{\lambda} \sum_j^n m_j^2/x_j}\delta\left(\sum_j x_j-1\right)	
\end{equation}
with $\lambda=\kappa^2$. Note that Equation \eqref{BdT-shift-general} coincides with Equation \eqref{BT-shift} for $n=2$. 

Strictly speaking, Equation \eqref{BT-shift} is only accurate for light hadrons in their ground states. However, as a first approximation, it is not unreasonable to assume the same correction for excited states. At the same time, this guarantees that the predicted Regge trajectories remain linear. Indeed, Equation \eqref{BT-shift}, together with Eqs. \eqref{MTM}, \eqref{MTB}, and \eqref{MTT}, have been used to fit the light hadron spectrum in Ref. \cite{Brodsky:2016rvj}, resulting in a universal $\kappa=0.523 \pm0.024$ GeV, with $m_{u/d}=0.046$ GeV and $m_s=0.357$ GeV. The use of Equation \eqref{BT-shift} has also been extrapolated to heavy quarks in order to predict the heavy-light and heavy-heavy hadron spectra in Refs. \cite{Dosch:2016zdv,Nielsen:2018ytt,Dosch:2015bca}, with the requirement that $\kappa$ is no longer universal in the heavy hadron sector. In an attempt to protect the universality of $\kappa$ in the heavy sector, Ref. \cite{Branz:2010ub} introduces another mass scale (instead of $\kappa$) in Equation \eqref{chi-BdT}.     

A theoretical shortcoming of Equation \eqref{BT-shift} is that, in the chiral limit, it predicts \cite{Li:2021jqb} $M_\pi^2 \propto 2m_{u/d}^2 (\ln\kappa^2/m_{u/d}^2 -\gamma_E)$ where $\gamma_E \approx 0.577216$ is Euler's constant, which is not consistent with the Gell-Mann-Oakes-Renner (GMOR) relation \cite{PhysRev.175.2195} which states that $M_\pi^2 \propto m_{u/d}$ in the chiral limit. An improved ansatz, consistent with the GMOR relation, has been proposed in Ref. \cite{Gutsche:2012ez}. Recently, this shortcoming has been more rigorously addressed in Refs. \cite{Li:2021jqb,DeTeramond:2021jnn} by  solving Equation \eqref{Mpara} with a longitudinal confining potential \cite{Sheckler:2020fbt,Li:2015zda}
\begin{equation}
	U_\parallel(x)=-\sigma^2 \partial_x (x(1-x)) \partial_x
\label{UL-BFLQ}
\end{equation}
where $\sigma$ is a mass scale. Reference \cite{Li:2021jqb} studies light mesons including their excited states while Ref. \cite{DeTeramond:2021jnn} focuses on the ground states of light and heavy mesons. 
While using Equation \eqref{UL-BFLQ}, Ref. \cite{DeTeramond:2021jnn} also discusses the relation of their approach to the 't Hooft equation. The idea to use the 't Hooft equation to go beyond the BdT prescription was first proposed in Ref. \cite{Chabysheva:2012fe}, with the unique goal of predicting the meson decay constants while leaving the predicted mass spectrum unchanged. The latter constraint is imposed by subtracting $M^2_\parallel$ from the 't Hooft potential. 

In this paper, we use the 't Hooft equation (without shifting the 't Hooft potential as in Ref. \cite{Chabysheva:2012fe}), together with the holographic Schr\"odinger equation, to compute the full hadron spectrum, i.e., the masses of the ground and excited states of mesons, baryons, and tetraquarks, including those with one or two heavy quarks. In Ref. \cite{Ahmady:2021lsh}, only the meson spectrum was studied. We shall also compare our results with the mass spectrum obtained using the alternative longitudinal potential given by Equation \eqref{UL-BFLQ}.

\section{The '\lowercase{t} Hooft Equation} 
The 't Hooft equation is derived from the QCD Lagrangian in $(1+1)$-dim and in the large $N_c$ approximation where only planar diagrams contribute. The result is \cite{tHooft:1974pnl}:
\begin{equation}
\left(\frac{m_q^2}{x}+\frac{m_{\bar{q}}^2}{1-x}\right)\chi(x) +\frac{g^2}{\pi} \mathcal{P} \int {\rm d}y \frac{|\chi(x)-\chi(y)|}{(x-y)^2}=M^2_\parallel \chi(x) \;,
  \label{tHooft}
\end{equation}
where $g=g_s \sqrt{N_c}$ is the (finite) 't Hooft coupling and $\mathcal{P}$ denotes the principal value prescription. Note that Equation \eqref{tHooft} can also be derived in the continuum limit of discretized $(1+1)$-dim light-front QCD \cite{Hornbostel:1988fb}. We use the 't Hooft equation for baryons and tetraquarks by making the transformation $\bar{q} \to [qq]$ for baryons, followed by $q \to [\bar{q}\bar{q}]$ for tetraquarks. Since the color interaction is invariant under these transformations, the longitudinal confinement scale, $g$, remains the same within a family of superpartners. The 't Hooft equation has been extensively studied in the literature \cite{Bars:1978,Zhitnitsky:1985um,Grinstein:1997xk,PhysRevLett.69.1018,Bergknoff:1976xr,Ji:2021,Lebed:2000,Ma:2021yqx,Umeeda:2021llf} and it is worth noting that, in the conformal limit, it possesses a gravity dual on $\mathrm{AdS}_3$ \cite{Katz:2007br}. Here, we only highlight how it is both complementary and consistent with the holographic Schr\"odinger equation

First, assuming a 't Hooft wave function of the form
\cite{DeTeramond:2021jnn}
\begin{equation}
	\chi(x) \approx x^{\beta_1}(1-x)^{\beta_2}
\end{equation}
then, near the endpoints, $x \to 0,1$,
\begin{equation}
	\frac{\pi m_i^2}{g^2}-1 + \pi \beta_i \cot(\pi \beta_i)=0 \;.
\label{Trans}
\end{equation}
Equation \eqref{Trans} implies that, in the chiral limit, $m_i \to 0$,  
\begin{equation}
\beta_i=\sqrt{3m_i^2/\pi g^2}	
\label{beta-chiral}
\end{equation}
and 
\begin{equation}
	M_\pi^2 = g\sqrt{\frac{\pi}{3}} (m_u + m_d) + \mathcal{O} (m_u + m_d)^2\,.
\label{GMOR-tHooft}
\end{equation}
Since, via Equation \eqref{MTM}, the holographic Schr\"odinger equation predicts a massless pion, it follows that the only contribution to the pion mass is generated by the 't Hooft equation.  Thus, together, the holographic Schr\"odinger equation and the 't Hooft equation correctly predicts the GMOR relation, $M_\pi^2 \sim m_{u/d}$. Notice also that in the chiral limit, $\beta_i \to 0$ and $\chi(x)=1$ as in light-front holography.  

Second, as already noted in Ref. \cite{Ahmady:2021lsh}, the 't Hooft potential is consistent with the holographic potential in that they both correspond to a linear confining potential in the non-relativistic limit, with $g =\kappa$ \cite{Ahmady:2021lsh}. To show this, first note that the 't Hooft potential in position space reads 
\begin{equation}
	U_\parallel^{\mathrm{tHooft}}(x^-)= \frac{g^2}{2} P^+|x^-|= g^2 P^+ b_\parallel
	\label{UL-x} \;,
\end{equation}
where $x^-=x^0-x^3$ is the light-front longitudinal distance, and $P^+$ is the light-front longitudinal momentum of the meson. The second equality follows from the fact that $x^+=x^0+x^3=0$, so that $|x^-|=2|x^3| \equiv 2 b_\parallel$.  Note that $\tilde{z}=P^+ x^-$ is the frame-independent longitudinal coordinate of Ref. \cite{Miller:2019ysh}.

Using the general relation between a frame-invariant light-front (LF) potential and a center of mass (CM) frame instant-form (IF) potential  \cite{Trawinski:2014msa}:
\begin{equation}
	U_{\mathrm{LF}}= V_{\mathrm{IF}}^2 + 4 m V_{\mathrm{IF}} \;,
	\label{LF-IF-potentials}
\end{equation}
it follows from Equation \eqref{LF-IF-potentials} that the (chiral-limit) quadratic light-front holographic potential, Equation \eqref{U-LFH}, corresponds to a linear instant form potential. The light-front 't Hooft potential, Equation \eqref{UL-x}, also corresponds to a linear instant form potential in the heavy quark (non-relativistic) limit where $m_Q \gg g$ and $P^+ \to 2m_Q$, where $m_Q$ is the heavy quark mass. In this non-relativistic limit, it is also appropriate to take $x \to 1/2$ in Equation \eqref{U-LFH}. Therefore, we see that, in the non-relativistic limit, the holographic potential and the 't Hooft potential are both equivalent to CM-frame linear instant-form potentials:
\begin{equation}
	V_\perp \to \frac{1}{2}\kappa^2 b_\perp\,,
\label{VperpIF}
\end{equation}
 and
\begin{equation}
	V_\parallel \to \frac{1}{2} g^2 b_\parallel \;.
\label{VparaIF}
\end{equation}
It then follows that $V_\perp=V_\parallel$, i.e., rotational symmetry is restored in the CM frame of a heavy-heavy meson if  $g=\kappa$.

Third, when solving the 't Hooft equation, an additional quantum number, which we shall label as $n_\parallel$, emerges. Thus, each hadronic state is identified by four quantum numbers, $n_\perp, L, S, n_\parallel$, with their squared masses given by 
\begin{equation}
	M_{M}^2=M^2_{\perp,M}(n_\perp, L_M, S_M, \kappa) + M^2_{\parallel,M}(n_\parallel, m_q, m_{\bar{q}}, g)\;,
	\label{MM}
\end{equation}
\begin{equation}
	M_{B}^2=M^2_{\perp,B}(n_\perp, L_B, S_D, \kappa) + M^2_{\parallel,B}(n_\parallel, m_q, m_{[qq]}, g) \;,
	\label{MB}
\end{equation}
\begin{equation}
	M_{T}^2= M^2_{\perp,B}(n_\perp, L_T, S_T, \kappa) + M^2_{\parallel,T}(n_\parallel, m_{[\bar{q}\bar{q}]}, m_{[qq]}, g)\,,	
	\label{MT}
\end{equation}
and their parity given by
\begin{equation}
	P=(-1)^{L_M+1}=(-1)^{L_B}=(-1)^{L_T}\;.
\label{P}
\end{equation}
In addition, where appropriate, the charge conjugation of mesons/tetraquarks is given by
\begin{equation}
	C=(-1)^{n_\parallel + L_M+S_M}=(-1)^{n_\parallel + L_T + S_T-1} \;.
\label{C}
\end{equation}
{\it \`A posteriori}, we find that 
\begin{equation}
	n_{\parallel} \ge n_{\perp} + L 
	\end{equation}
i.e., in any hadron, an orbital and/or radial excitation in the transverse dynamics is always accompanied by an excitation in the longitudinal dynamics. This is a signature of the underlying link between the holographic Schr\"odinger equation and the 't Hooft equation. This is perhaps not surprising since they both derive from the QCD Lagrangian, albeit in different dimensions and limits.

Fourth, the 't Hooft equation correctly predicts the scaling law of the decay constant of heavy-light mesons with the heavy quark mass: $f_M \propto 1/\sqrt{m_Q}$ \cite{Grinstein:1997xk}. Since the decay constant depends on the meson wave function at zero transverse separation, it follows that this remains true if the meson wave function is given by Equation \eqref{full-mesonwf}.

Unlike the holographic Schr\"odinger equation, the 't Hooft equation does not have exact analytical solutions. We shall solve it numerically using the matrix method outlined in Ref. \cite{Chabysheva:2012fe}, and already used in Ref. \cite{Ahmady:2021lsh}.

\begin{table}[httb]
 \begin{tabular}{|  l  | c |  c  |  c  | c  | c | c |}
 \cline{1-6}
 Hadron & $g$ & $m_{u/d}$ & $m_s$ & $m_c$ & $m_b$ \\
 \cline{1-6} 
 Light~ & 0.128 & 0.046 & 0.357 & - & - \\
 Heavy-light & 0.410 & 0.330 & 0.500 & 1.370 & 4.640 \\
 Heavy-heavy~ & 0.523  & - & - & 1.370 & 4.640 \\
 \cline{1-6} 
\end{tabular}
\caption{Quark masses and longitudinal confinement scale, $g$, in $\mathrm{GeV}$ used in conjunction with the 't Hooft potential. Note that we use $\kappa=0.523$ GeV for all hadrons, and that it coincides with $g$ for hadrons with two heavy quarks.}
\label{Tab:params}
\end{table}



\section{Computing the hadron spectrum}
We now use Eqs. \eqref{MM}, \eqref{MB}, \eqref{MT}, \eqref{P} and \eqref{C} to compute the hadron spectrum with the quark masses and values of $g$ shown in Table \ref{Tab:params}, together with $\kappa =0.523$ GeV for all hadrons. In comparison to Ref. \cite{Ahmady:2021lsh}, we allow the light quark masses to vary between light and heavy-light hadrons, with the goal of achieving a better global description of the full hadron spectrum. We adopt the simplest assumption that the (anti)diquark mass is the sum of the (anti)quark masses, i.e., the (anti)diquark is essentially a cluster of two (anti)quarks. Note that we identify the superpartners as in Ref. \cite{Nielsen:2018uyn}. Our results are shown in Table \ref{Tab:light} for the light hadrons, in Table \ref{Tab:heavy-light} for hadrons with one heavy quark, and in Table \ref{Tab:heavy} for hadrons with two heavy quarks.

In general, our meson and baryon masses are in good agreement with the data, with a discrepancy not exceeding $13\%$, except for $\eta^\prime(958)$, where it is $21\%$. For the tetraquark candidates, the discrepancy does not exceed $18\%$, with notable exceptions for $f_0(500)$ and $f_0(980)(a_0(980))$, where the discrepancies are very large. The agreement for mesons and baryons is thus more impressive than for tetraquarks. In fact, the large discrepancies for the tetraquark candidates, $f_0(500)$ and $f_0(980)(a_0(980))$, are already present before any supersymmetry breaking by longitudinal dynamics. This can be seen by comparing the values of $M_\perp$ to the physical masses of $f_0(500)$ and $f_0(980)(a_0(980))$ in Table \ref{Tab:light}. Thus these discrepancies are mostly inherited from the original formulation of supersymmetric light-front holography, and are not alleviated by longitudinal dynamics.

It is instructive to compare our results with those obtained using the alternative longitudinal potential given by Equation \eqref{UL-BFLQ}. In this case, an analytical formula can be derived for the meson mass spectrum \cite{Li:2021jqb}:
\begin{equation}
	M_{\parallel,M}^2= \sigma (m_q + m_{\bar{q}}) (2n_\parallel +1) + \sigma^2 n_\parallel (n_\parallel + 1) + (m_q + m_{\bar{q}})^2 \;.
\label{Mparallel-BLFQ}
\end{equation}
Reference \cite{DeTeramond:2021jnn} shows that the restoration of rotational symmetry in heavy-heavy mesons implies the constraint: 
\begin{equation}
	\sigma=\frac{\kappa^2}{m_Q + m_{\bar{Q}}} \;,
\label{non-rel-constraint}
\end{equation}
which is to be contrasted with the analogous constraint, $g=\kappa$, for the 't Hooft potential: see Equation \eqref{VperpIF} and Equation \eqref{VparaIF}. To compute the hadron masses, we proceed in the same way as with the 't Hooft potential: within a family of superpartners, we use Equation \eqref{Mparallel-BLFQ}, with $\bar{q} \to [qq]$ (for the baryon), followed  by $q \to [\bar{q}\bar{q}]$ (for the tetraquark). Again, we assume that the (anti)diquark mass is twice the quark mass. This means that our goal is only to investigate the sensitivity of our results to the analytical form of the longitudinal potential. As can be seen in Tables \ref{Tab:light-BLFQ}, \ref{Tab:heavy-light-BLFQ} and \ref{Tab:heavy-BLFQ},  the agreement with data is similar in quality to that achieved with the 't Hooft potential, provided that $\sigma$ remains universal while  $\kappa$ is allowed to vary across the spectrum, subject to the constraint given by Equation \eqref{non-rel-constraint}. Their numerical values, as well as those of the quark masses, are given in  Table \ref{Tab:params-bflq}.  

\begin{table}[httb]
 \begin{tabular}{|  l   c |  c |  c  |  c  | c  | c |}
 \cline{1-7}
Hadron &  & $\kappa$ & $m_{u/d}$ & $m_s$ & $m_c$ & $m_b$ \\ \cline{1-6}
 \cline{1-7} 
 Light~ & & 0.523 & 0.046  & 0.357 & - & - \\ \cline{1-7}
 \multirow{2}{*}{Heavy-light} & 1 $c$-quark & 0.580 & 0.330 & 0.500 & 1.470 & - \\ 
 & 1 $b$-quark & 1.000 & 0.330 &  0.500 & - & 4.670 \\ \cline{1-7}
\multirow{2}{*}{Heavy-heavy} & 2 $c$-quarks  & 0.600 & - & - & 1.470 & - \\
& 2 $b$-quarks & 1.070 & - & - & - & 4.670 \\
 \cline{1-7} 
\end{tabular}
\caption{Quark masses and the transverse confinement scale, $\kappa$, in $\mathrm{GeV}$ used in conjunction with the alternative longitudinal potential, Equation \eqref{UL-BFLQ}. Note that we use $\sigma=0.12$ GeV for all hadrons, and that the constraint, Equation \eqref{non-rel-constraint}, is satisfied for hadrons with two heavy quarks.}
\label{Tab:params-bflq}
\end{table}

Interestingly, it seems that either the longitudinal or the transverse confinement scale can be  universal across the spectrum, depending on whether we choose Equation \eqref{UL-BFLQ} or the 't Hooft potential to model the longitudinal dynamics. The 't Hooft potential offers the advantage that its connection to the QCD Lagrangian is known. For this reason, it remains our focus in this paper.

\begin{figure}[http]
		\includegraphics[scale=0.6]{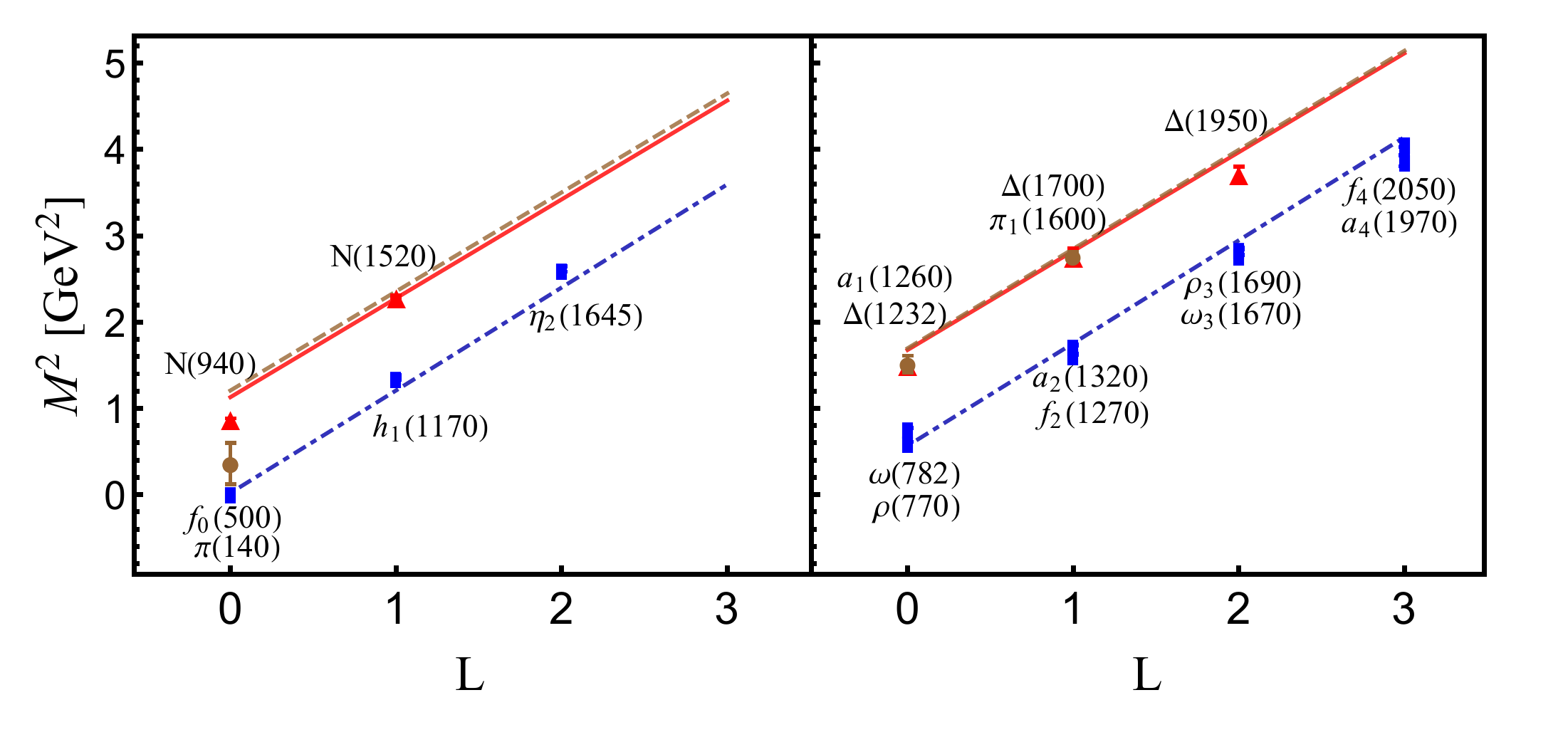}
		\caption{Regge trajectories for the supersymmetric meson-baryon-tetraquark partners for the unflavored light-light sector. The results are compared to the PDG data \cite{Zyla:2020zbs}. The mesons, baryons and tetraquarks trajectories are the dot-dashed-blue, solid-red, and dashed-brown lines, respectively. For the left panels, $S=0$ and for the right panels, $S=1$.}
\label{Regge-light}
\end{figure}

\begin{figure}[http]
	\includegraphics[scale=0.6]{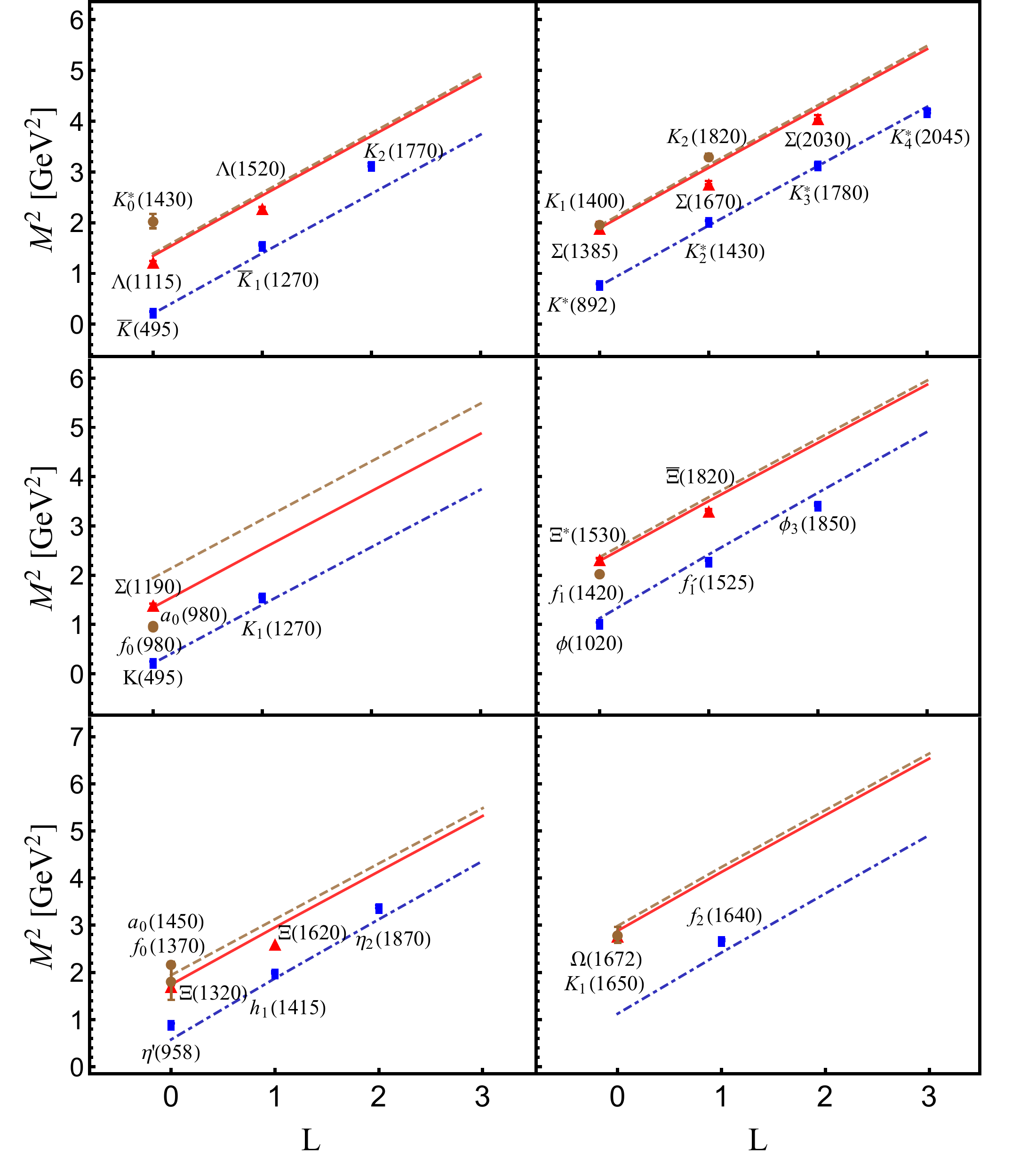}
	\caption{Regge trajectories for the supersymmetric meson-baryon-tetraquark partners for the light-light sector with at least one strange quark. The results are compared to the PDG data \cite{Zyla:2020zbs}. The mesons, baryons, and tetraquarks trajectories are the dot-dashed-blue, solid-red, and dashed-brown lines, respectively. For the left panels, $S=0$ and for the right panels, $S=1$.}
\label{Regge-light-strange}
\end{figure}

\begin{figure}[http]
	\includegraphics[scale=0.6]{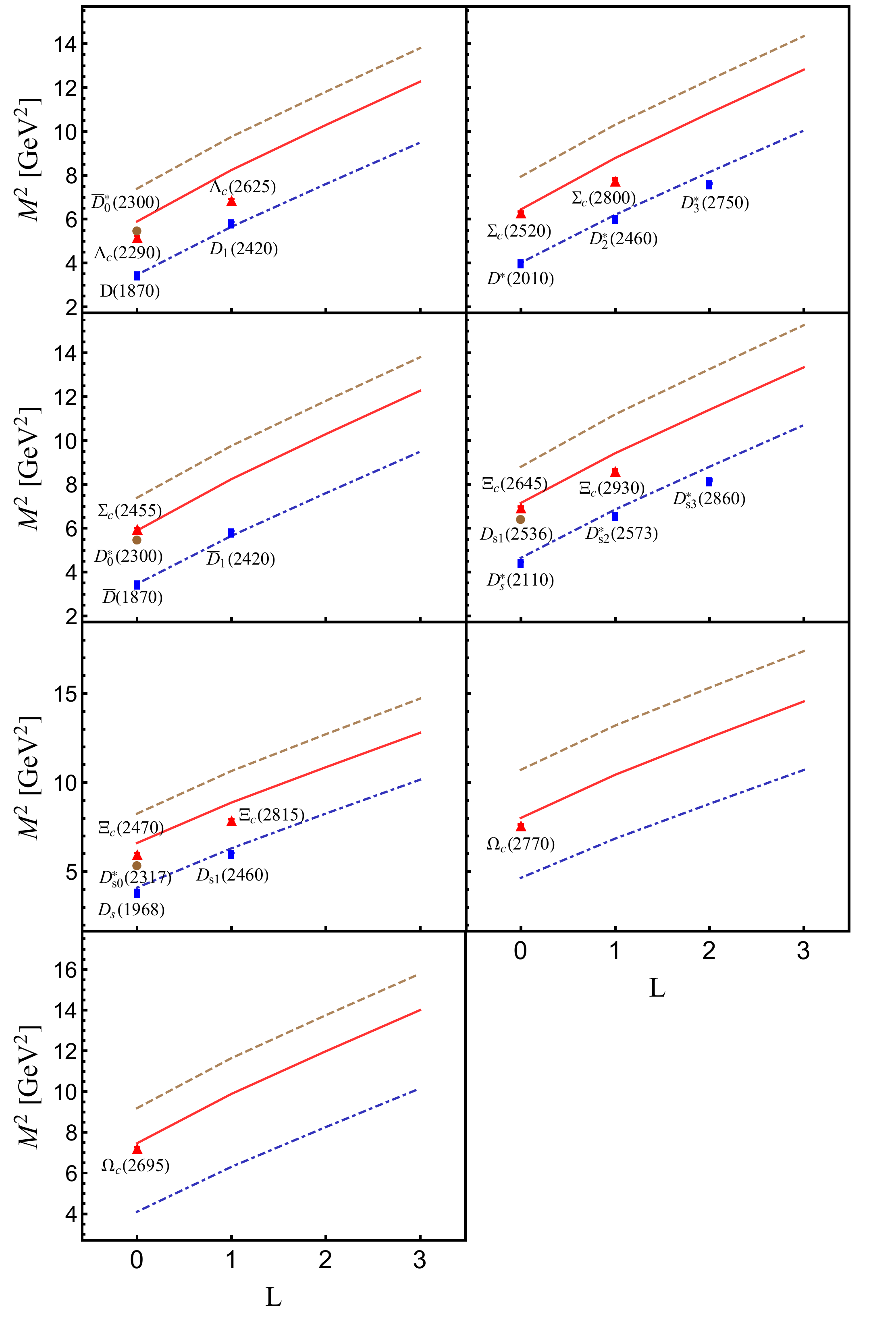}
	\caption{Regge trajectories for the supersymmetric meson-baryon-tetraquark partners for the heavy-light sector with one charm quark. The results are compared to the PDG data \cite{Zyla:2020zbs}. The mesons, baryons, and tetraquarks trajectories are indicated by dot-dashed-blue, solid-red, and dashed-brown lines, respectively. For the left panels, $S=0$ and for the right panels, $S=1$.}
\label{Regge-heavy-light-d}
\end{figure}

\begin{figure}[http]
	\includegraphics[scale=0.6]{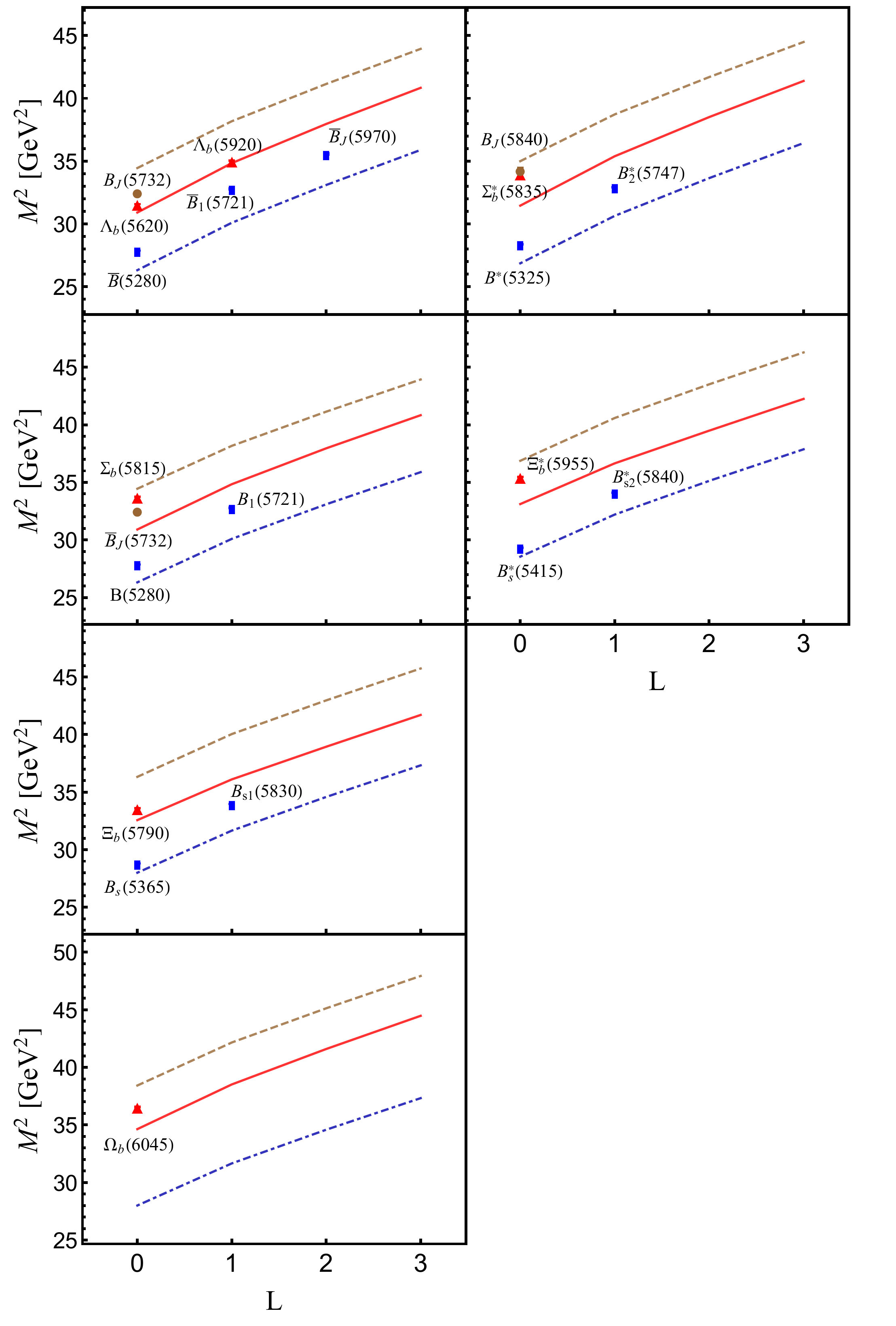}
	\caption{Regge trajectories for the supersymmetric meson-baryon-tetraquark partners for the heavy-light sector with one bottom quark. The results are compared to the PDG data \cite{Zyla:2020zbs}. The mesons, baryons, and tetraquarks trajectories are indicated by dot-dashed-blue, solid-red, and dashed-brown lines, respectively. For the left panels, $S=0$ and for the right panels, $S=1$.}
\label{Regge-heavy-light-b}
\end{figure}

\begin{figure}[http]
	\includegraphics[scale=0.6]{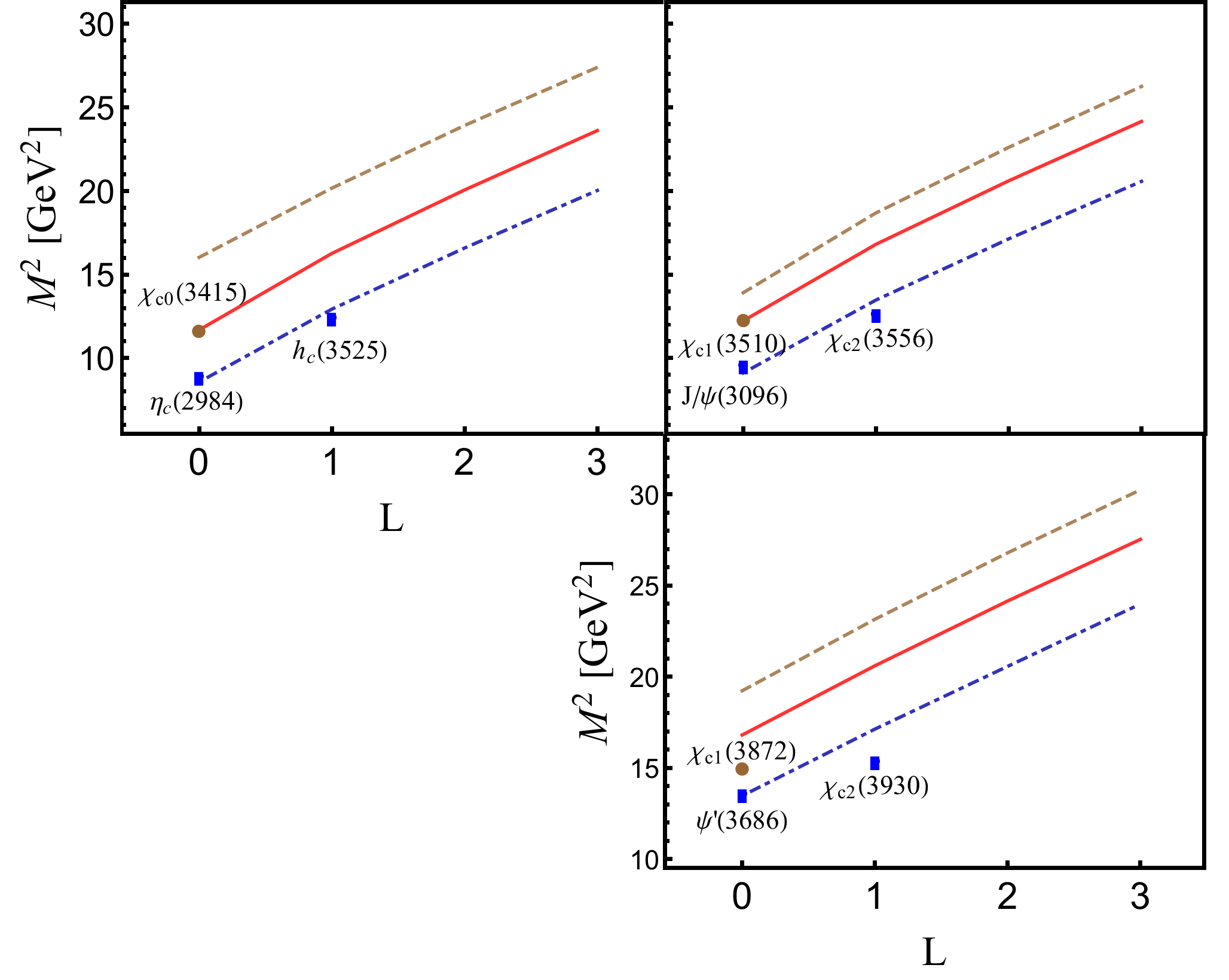}
	\caption{Regge trajectories for the supersymmetric meson-baryon-tetraquark partners for the doubly charmed (heavy-heavy) sector. The results are compared to the PDG data \cite{Zyla:2020zbs}. The mesons, baryons, and tetraquarks trajectories are indicated by dot-dashed-blue, solid-red, and dashed-brown lines, respectively. For the left panels, $S=0$ and for the right panels, $S=1$.}
\label{Regge-heavy-cc}
\end{figure}

\begin{figure}[http]
	\includegraphics[scale=0.6]{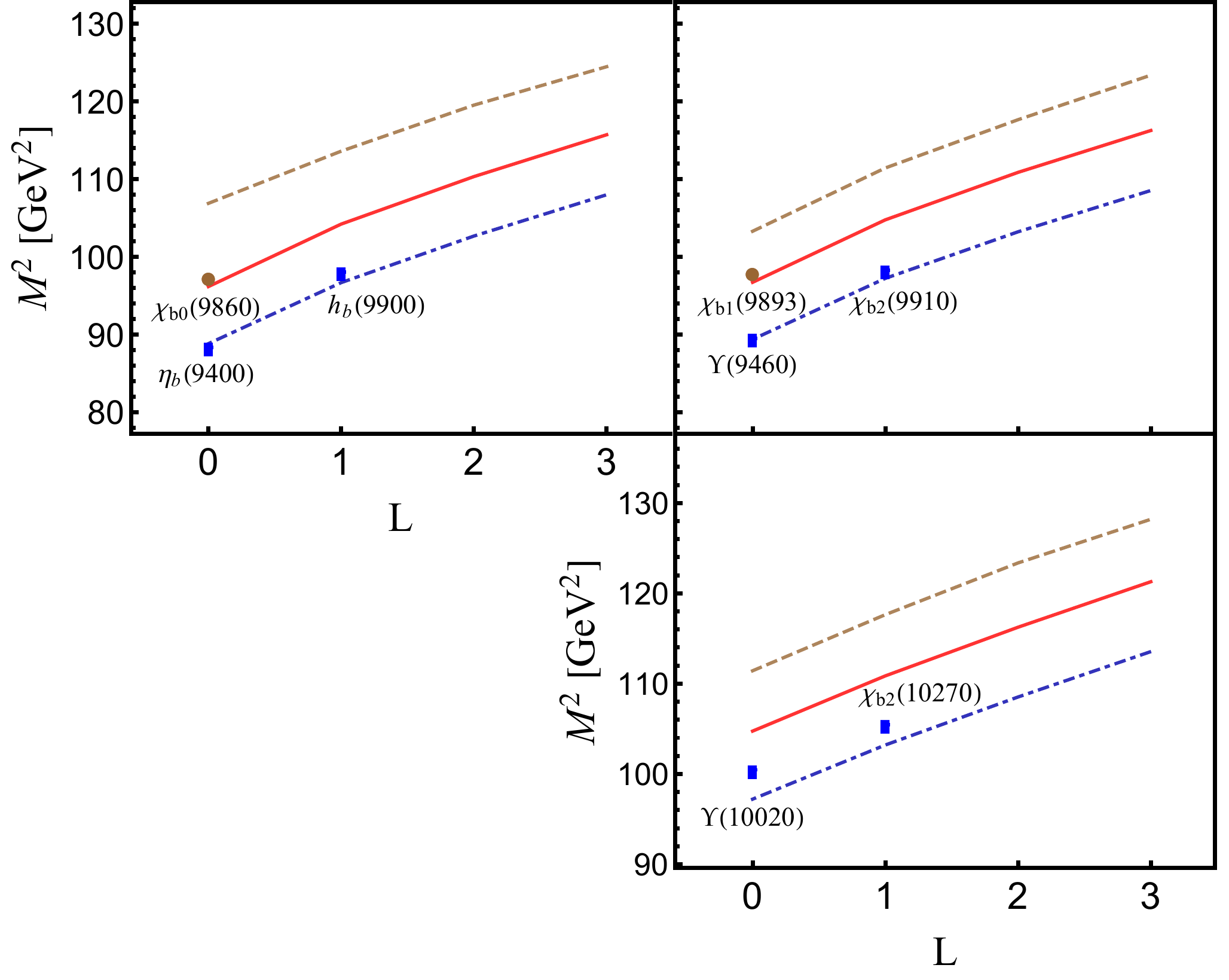}
	\caption{Regge trajectories for the supersymmetric meson-baryon-tetraquark partners for the doubly bottom (heavy-heavy) sector. The results are compared to the PDG data \cite{Zyla:2020zbs}. The mesons, baryons, and tetraquarks trajectories are indicated by dot-dashed-blue, solid-red, and dashed-brown lines, respectively. For the left panels, $S=0$ and for the right panels, $S=1$.}
\label{Regge-heavy-bb}
\end{figure}

\begin{figure}[http]
		\includegraphics[scale=0.7]{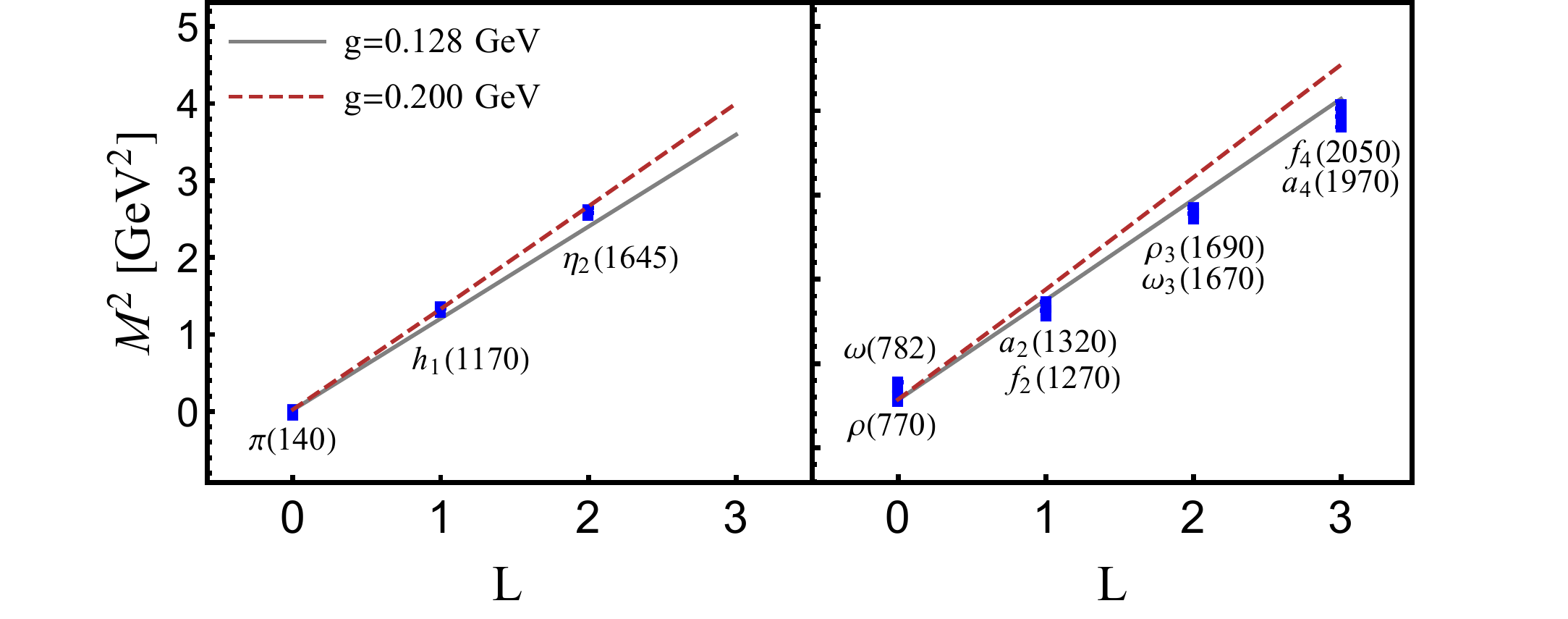}
		\caption{Regge trajectories for $\pi$ and $\rho$ families using different values of $g$, e.g., $g=0.128$ GeV (solid-gray) and $g=0.200$ GeV (dashed-dark-red).}
\label{diff-g-light}
\end{figure}

\begin{figure}[http]
		\includegraphics[scale=0.7]{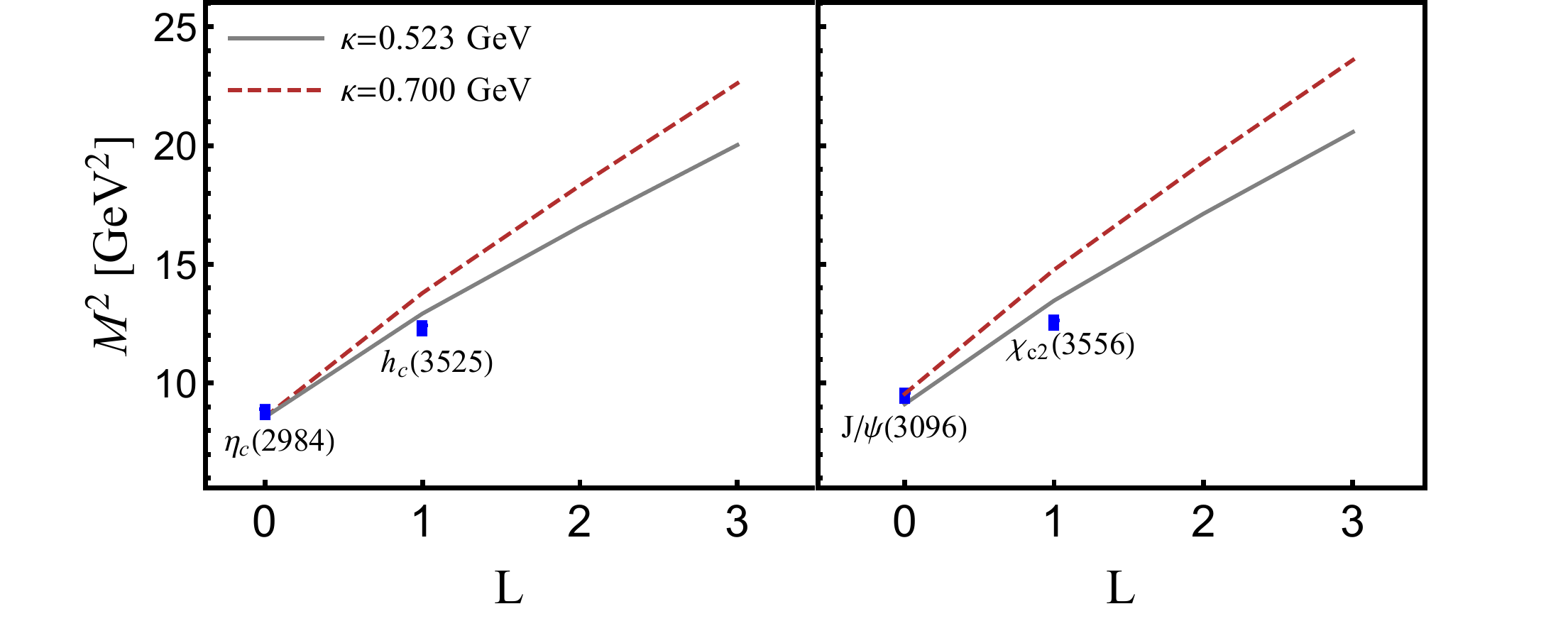}
		\caption{Regge trajectories for $\eta_c$ and $J/\psi$ families using different values of $\kappa$, e.g., $\kappa=0.523$ GeV (solid-gray) and $\kappa=0.700$ GeV (dashed-dark-red).}
\label{diff-kappa-heavy}
\end{figure}

We now proceed to show the Regge trajectories obtained using the 't Hooft potential: Figs. \ref{Regge-light} and \ref{Regge-light-strange} for the light hadrons, Figs. \ref{Regge-heavy-light-d} and \ref{Regge-heavy-light-b} for hadrons with one heavy quark, and Figs. \ref{Regge-heavy-cc} and \ref{Regge-heavy-bb} for hadrons with two heavy quarks. As expected, the agreement is best for mesons and baryons. Except for the lowest-lying pseudoscalar mesons for which $M_\perp=0$, $M_\perp \gg M_\parallel$ for light hadrons. On the other hand, $M_\parallel \gg M_\perp$ for hadrons with one or two heavy quarks. It is therefore legitimate to investigate the sensitivity of our Regge trajectories to a small variation in $g$ (for light hadrons) or $\kappa$ (for heavy hadrons), while keeping the quark masses fixed. In Fig. \ref{diff-g-light}, we show the effect of changing $g$ from $0.128$ to $0.200$ GeV on the $\pi$ and $\rho$ Regge trajectories. As can be seen, the slopes of the trajectories are sensitive to this variation in $g$. Note that, as expected, the pion mass also changes from $0.140$ to $0.163$ GeV but this is not visible on the mass scale of Fig. \ref{diff-g-light}. Meanwhile, in Fig. \ref{diff-kappa-heavy}, we illustrate the effect of changing $\kappa$ from $0.523$ to $0.700$ GeV on the $\eta_c$ and $J/\psi$ Regge trajectories. We can see that the Regge slopes are significantly changed with this variation in $\kappa$. These observations are also typical for the other Regge trajectories. Therefore, the precise slopes and locations of Regge trajectories, for a given set of quark masses, are sensitive to both $g$ and $\kappa$. It is worth highlighting that, in our approach, $\kappa$ remains universal across the full hadron spectrum.

\begin{figure}[http]
		\includegraphics[scale=0.6]{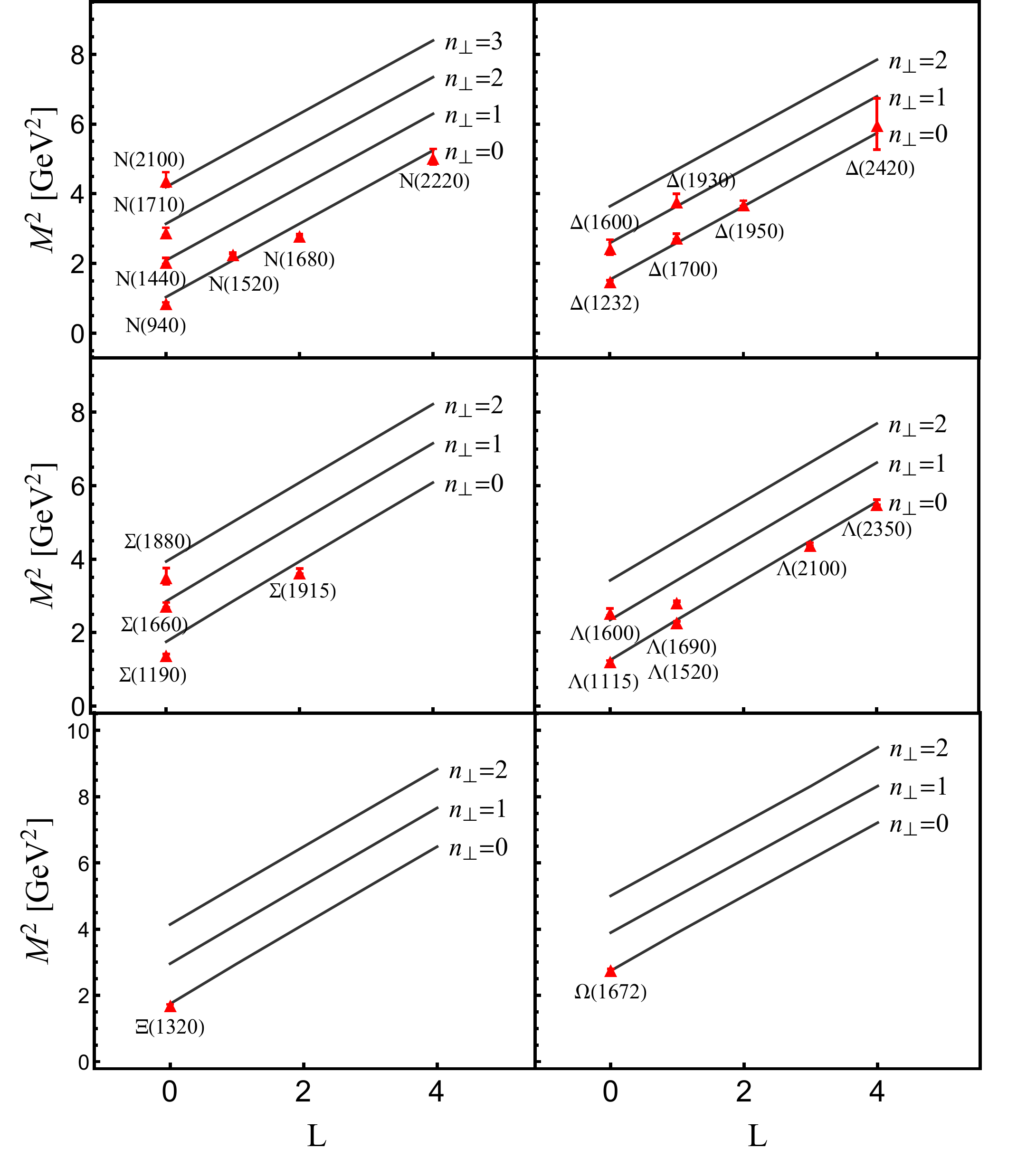}
		\caption{Regge trajectories for the light baryons, compared to the PDG data \cite{Zyla:2020zbs}.}
\label{baryons-light-light}
\end{figure}

\begin{figure}[http]
	\includegraphics[scale=0.57]{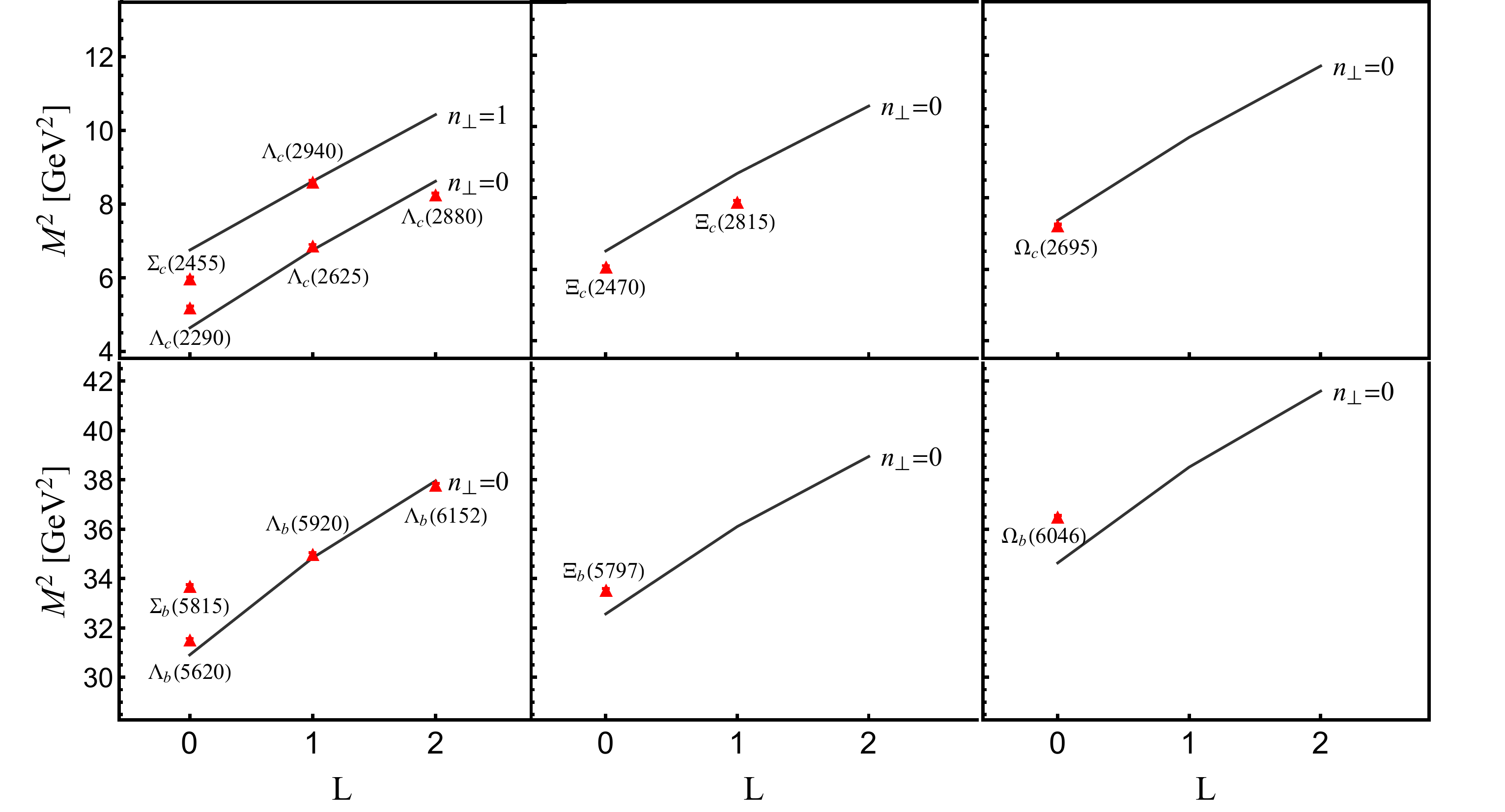}
	\caption{Regge trajectories for the heavy baryons involving one charm (bottom) quark, compared to the PDG data \cite{Zyla:2020zbs}.}
\label{baryons-heavy-light}
\end{figure}

In Figs. \ref{baryons-light-light} and \ref{baryons-heavy-light}, we compare our results for all baryons, including those with no identified superpartners from the PDG. Using $\kappa=0.500\pm 0.024$ GeV, we achieve very good agreement with the data.

\section{Conclusions}
We have shown that the 't Hooft equation is complementary to the holographic Schr\"odinger equation in predicting the full hadron spectrum. While the holographic Schr\"odinger equation generates the hadronic mass in the chiral limit of QCD with a universal emerging transverse confinement scale, the 't Hooft equation generates the contribution to the hadronic mass due to nonzero quark masses and longitudinal confinement. Agreement with spectroscopic data spectra is generally good, while the disagreement for the tetraquark candidates, $f_0(500)$ and $f_0(980)(a_0(980))$, already present when neglecting longitudinal dynamics, persists.   

\section{Acknowledgements}
We thank Stan Brodsky and Guy de T\'eramond for useful discussions. R.S. and M.A. are supported by individual Discovery Grants No. SAPIN-2020-00051 and No. SAPIN-2021-00038 from the Natural Sciences and Engineering Research Council of Canada (NSERC). C. M. is supported by new faculty start up funding by the Institute of Modern Physics, Chinese Academy of Sciences, Grant No. E129952YR0. C.M. also thanks the Chinese Academy of Sciences President's International Fellowship Initiative for the support via Grant No. 2021PM0023.

\bibliographystyle{apsrev}
\bibliography{ref}

	\begin{table}[hbt!]
		\centering
		\resizebox{1\textwidth}{!}{
			\begin{tabular}{|c c c c c |c c c c c |c c c c c|}
				\hline
				\multicolumn{5}{|c}{Meson} & \multicolumn{5}{|c|}{Baryon} & \multicolumn{5}{c|}{Tetraquark}\\
				$J^{P(C)}$ & Name & $M_{\parallel}$ & $M_\perp$ & $M$ &  $J^{P(C)}$ & Name & $M_{\parallel}$ & $M_\perp$  & $M$ & $J^{P(C)}$ & Name & $M_{\parallel}$ & $M_\perp$  & $M$ \\
				\hline
				$0^{-+}$ & $\pi(140)$ & 140 & 0 & 140  &- & - & - & - & - &-&- &- &-&-\\
				$1^{+-}$ & $h_{1}(1170)$ & 335  & 1046 & 1098 & $(1/2)^+$ & $N(940)$ & 188 & 1046 & 1063 &  $0^{++}$ & $f_0(500)$ & 335 &  1046 & 1098 \\
				$2^{-+}$ &$\eta_{2}(1645)$ & 460 & 1479 & 1549 &  $(3/2)^-$ & $N(1520)$ & 296  & 1479 & 1508 & $1^{-+}$ & - & 408 &  1479 & 1534  \\
				\hline
				$1^{--}$	&$\rho(770)$, $\omega(780)$ & 140 & 740 & 753 & - & - & -& - &-&-&-&-&-&-\\
				$2^{++}$ &$a_{2}(1320)$, $f_{2}(1270)$ & 335 &  1281 & 1324 & $(3/2)^+$ & $\Delta(1232)$ & 188 & 1281 & 1295 &  $1^{++}$ & $a_{1}(1260)$ & 235 & 1281 & 1302 \\
				$3^{--}$ & $\rho_{3}(1690)$, $\omega_{3}(1670)$ & 460 & 1654 & 1717 & $(3/2)^-$ & $\Delta(1700)$ & 296 & 1654 & 1680 &  $1^{-+}$ & $\pi_{1}(1600)$ & 335 &  1654 & 1688  \\
				$4^{++}$ & $a_{4}(1970)$, $f_{4}(2050)$ & 559 & 1957 & 2035 & $(7/2)^+$ &$\Delta(1950)$ & 372  & 1957 & 1992 & - & - & - & - & -\\
				\hline
				$0^-$	& $\bar{K}(495)$ & 456 & 0 & 456 &- & - & - & - & - & - & - & -&-&-\\
				$1^+$&	$\bar{K}_{1}(1270)$ & 550 & 1046 & 1182 & $(1/2)^+$ & $\Lambda(1115)$ & 500 &  1046 & 1159 & $0^+$  &  $K^{*}_{0}(1430)$ & 546 & 1046 & 1180 \\
				$2^-$	& $K_{2}(1770)$ & 617 & 1479 & 1603 & $(3/2)^-$ & $\Lambda(1520)$ & 588 & 1479 & 1592 & $1^-$ & - & 633 &  1479 & 1609 \\
				\hline
				$0^-$	& $\bar{K}(495)$ & 456 & 0 & 456 & - &  - &-  & - & - & - & - & -&-&-\\
				$1^+$ &	$\bar{K}_{1}(1270)$ & 550 & 1046 & 1182 & $(1/2)^+$ & $\Sigma(1190)$ & 500 & 1046 & 1159 & $0^{++}$ & $a_{0}(980)$ & 920 &  1046 & 1393 \\
				&  &   & & & & & & & &  & $f_{0}(980)$ & & & \\
				\hline
				$1^-$	& $K^{*}(890)$ & 456  & 740  & 869 &   - & -& -&  - & -&- & -& -&-&- \\
				$2^+$ & $K^{*}_{2}(1430)$ & 550 & 1281 & 1394 & $(3/2)^+$ & $\Sigma(1385)$ & 500 & 1281 & 1375  & $1^+$ & $K_{1}(1400)$ & 546 &  1281 & 1392  \\
				$3^-$	& $K^{*}_{3}(1780)$ & 617  & 1654 & 1765  & $(3/2)^-$ & $\Sigma(1670)$ & 588  & 1654 & 1755 & $2^-$ & $K_{2}(1820)$ & 633 & 1654 & 1771 \\
				$4^+$ &	$K^{*}_{4}(2045)$ & 672 & 1957 & 2069 & $(7/2)^+$ & $\Sigma(2030)$ & 650  & 1957 & 2062  & - & - & -&-&-\\
				\hline
				$0^{-+}$ & $\eta^\prime(958)$ & 759 & 0 & 759 & - &- &- &-& -&-&-&-&-&- \\
				$1^{+-}$ & $h_{1}(1380)$ & 883 & 1046 & 1369 & $(1/2)^+$ & $\Xi(1320)$ & 805  & 1046 & 1320  & $0^{++}$ & $f_{0}(1370)$ & 920 & 1046 & 1393 \\
				& & &  & & & & & & & & $a_{0}(1450)$ & & & \\
				$2^{-+}$&	$\eta_{2}(1870)$ & 968 & 1479 & 1768 &  $(3/2)^-$ & $\Xi(1620)$ & 876  & 1479 & 1719  & $1^{-+}$ & - & 969 & 1479 & 1768 \\
				\hline
				$1^{--}$ & $\Phi(1020)$ & 759 & 740 & 1060  & - & - & - & -&-&-&-&-&-&-\\
				$2^{++}$	& $f'_{2}(1525)$ & 883 &  1281 & 1556 &  $(3/2)^+$ & $\Xi^{*}(1530)$ & 805 & 1281 & 1513 &  $1^{++}$ & $f_{1}(1420)$ & 850  & 1281 & 1537 \\
				
				& &	&  & & &  & & & & & $a_{1}(1420)$ & & & \\
				$3^{--}$&	$\Phi_{3}(1850)$ & 968 & 1654 & 1916 & $(3/2)^-$ & $\Xi(1820)$ & 876  & 1654 & 1872 &  - & - &- & - & -\\
				\hline
				$2^{++}$ &	$f_{2}(1640)$ & 883  & 1281 & 1556 &  $(3/2)^+$ &$\Omega(1672)$ & 1114 & 1281 & 1698  & $1^+$ & $K_{1}(1650)$ & 1159  & 1281 & 1728  \\
				\hline
		\end{tabular}}
\caption{Computed masses (in MeV), using the 't Hooft potential, for the light hadrons compared to the PDG data \cite{Zyla:2020zbs}.}
\label{Tab:light}
\end{table}

	\begin{table}[hbt!]
	\centering
	\resizebox{0.9\textwidth}{!}{
		\begin{tabular}{|c c c c c|c c c c c|c c c c c|}
			\hline
			\multicolumn{5}{|c}{Meson} & \multicolumn{5}{|c|}{Baryon} & \multicolumn{5}{c|}{Tetraquark}\\
			$J^{P(C)}$	& Name & $M_{\parallel}$ &  $M_\perp$  &  $M$ &  $J^{P(C)}$	& Name & $M_{\parallel}$ & $M_\perp$ & $M$  &	$J^{P(C)}$	& Name & $M_{\parallel}$ & $M_\perp$ &  $M$ \\
			\hline
			$0^-$ &	$D(1870)$ & 1861 & 0  & 1861 &  - & - &  - & - & - &-&-&- &-&-\\
			$1^+$ &	$D_{1}(2420)$ & 2135  & 1046  & 2377  & $(1/2)^+$ & $\Lambda_{c}(2290)$ & 2191  & 1046  & 2428 & $0^+$ & $\bar{D^{*}_{0}}(2400)$ & 2510 & 1046 & 2719  \\
			$2^-$ &	$D_{J}(2600)$ & 2326  & 1479 & 2756 & $(3/2)^-$ & $\Lambda_{c}(2625)$ & 2460  & 1479 & 2870 & $1^-$ & - & 2751  & 1479 & 3123  \\
			\hline
			$0^-$ & $\bar{D}(1870)$ & 1861  & 0 & 1861  & - & - & - & - & - & -&-&-&-&-\\
			$1^+$ & $\bar{D}_{1}(2420)$ & 2135  & 1046  & 2377 &  $(1/2)^+$ & $\Sigma_{c}(2455)$ & 2191  & 1046 & 2428 & $0^+$ & $D^{*}_{0}(2400)$ & 2510  & 1046  & 2719  \\
			\hline
			$1^-$ & $D^{*}(2010)$ & 1861  & 740  & 2003  & - & - & - & - &-&-&-&-&-&-\\
			$2^+$ & $D^{*}_{2}(2460)$ & 2135  & 1281 & 2490 &  $(3/2)^+$ & $\Sigma^{*}_{c}(2520)$ & 2191  & 1281 & 2538 & $1^+$ & $D(2550)$ & 2510 & 1281  & 2818\\
			$3^-$ & $D^{*}_{3}(2750)$ & 2326  & 1654 & 2854 &  $(3/2)^-$ &$\Sigma_{c}(2800)$ & 2460  & 1654 & 2964 & - &-&-&-&-\\
			\hline
			$0^-$ & $D_{s}(1968)$ & 2025 & 0 & 2025 &- &- &- &-&-&-&-&-&-&-\\
			$1^+$ & $D_{s1}(2460)$ & 2283& 1046 & 2511 & $(1/2)^+$ & $\Xi_{c}(2470)$ & 2348 & 1046 & 2570 & $0^+$ & $\bar{D}^{*}_{s0}(2317)$ & 2676  & 1046  & 2873  \\
			$2^-$ & $D_{s2}?$ & 2464 & 1479  & 2874  & $(3/2)^-$ & $\Xi_{c}(2815)$ & 2586  & 1479  & 2979 & $1^-$ &-& 2908   & 1479 & 3262 \\
			\hline
			$1^-$ &	$D^{*}_{s}(2110)$ & 2025 & 740 & 2156 & - &  - &- &-&-&-&-&-&-&-\\
			$2^+$ & $D^{*}_{s2}(2573)$ & 2283  & 1281 & 2618 & $(3/2)^+$ & $\Xi^{*}_{c}(2645)$ & 2348  & 1281 & 2675  & $1^+$ &$D_{s1}(2536)$ & 2676  & 1281 & 2967 \\
			$3^-$ & $D^{*}_{s3}(2860)$ & 2464  & 1654  & 2968 & - &- &- & -&-&-&-&-&-&-\\
			\hline
			$1^+$ & $\bar{D}_{s1}?$ & 2283  & 1046 & 2511  & $(1/2)^+$ &$\Omega_{c}(2695)$ & 2524  & 1046 & 2732 & $0^+$ & - & 2845  & 1046  & 3031  \\
			\hline
			$2^+$ & $D^{*}_{s2}?$ & 2283 & 1281 & 2618 & $(3/2)^+$ & $\Omega_{c}(2770)$ & 2524 & 1281  & 2830 & $1^+$ & - & 3012 & 1281 & 3273  \\
			\hline \hline
			$0^-$ & $\bar{B}(5280)$ & 5130  & 0  & 5130 &-&- & -& -& -&-&-&-&-&-\\
			$1^+$ & $\bar{B}_{1}(5720)$ & 5385 & 1046  & 5486 & $(1/2)^+$ & $\Lambda_{b}(5620)$ & 5460 & 1046 & 5559   & $0^+$ & $B_{J}(5732)$ & 5775  & 1046  & 5869  \\
			$2^-$ & $\bar{B}_{J}(5970)$ & 5560 & 1479 & 5753 & $(3/2)^-$ & $\Lambda_{b}(5920)$ & 5714  & 1479  & 5902  & $1^-$ &-& 5999  & 1479  & 6179 \\
			\hline
			$0^-$ &$B(5280)$ & 5130 & 0 & 5130 &- &- &- &- &-&-&-&-&-&- \\
			$1^+$ & $B_{1}(5720)$ & 5385 & 1046 & 5486  & $(1/2)^+$ & $\Sigma_{b}(5815)$ & 5460  &  1046  & 5559 & $0^+$ &$\bar{B}_{J}(5732)$ & 5775 & 1046 & 5869 \\
			\hline
			$1^-$ & $B^{*}(5325)$ & 5130  & 740 & 5183 & - &- &- &-&-&-&-&-&-&-\\
			$2^+$	& $B^{*}_{2}(5747)$ & 5385& 1281  & 5535 &  $(3/2)^+$ &$\Sigma^{*}_{b}(5835)$ & 5460  & 1281 & 5608  & $1^+$ &$B_{J}(5840)$ &  5775  & 1281& 5915 \\
			\hline
			$0^-$ & $B_{s}(5365)$ & 5292 & 0 & 5292 & - & - & - & - & - & -&-&-&-&-\\
			$1^+$	& $B_{s1}(5830)$ & 5528  & 1046  & 5626 &  $(1/2)^+$ & $\Xi_{b}(5790)$ & 5610& 1046 & 5707  & $0^+$ &$\bar{B}^{*}_{s0}?$ & 5936  & 1046 & 6027  \\
			\hline
			$1^-$	& $B^{*}_{s}(5415)$ & 5292  & 740 & 5343 & - & - & - &-&-&-&-&-&-&-\\
			$2^+$ &	$B^{*}_{s2}(5840)$ & 5528 & 1281 &  5674 &  $(3/2)^+$ & $\Xi^{*}_{b}(5950)$ &5610 & 1281& 5754 & $1^+$ & $B_{s1}?$ & 5936  & 1281 & 6073 \\
			\hline
			$1^+$& $B_{s1}?$ & 5528 & 1046 & 5626  & $(1/2)^+$ & $\Omega_{b}(6045)$ & 5791  & 1046 & 5885 & $0^+$ &-& 6110  & 1046 &  6199 \\
			\hline
	\end{tabular}}
\caption{Computed masses (in MeV), using the 't Hooft potential, for hadrons with one heavy quark, compared to the PDG data \cite{Zyla:2020zbs}.}
\label{Tab:heavy-light}
\end{table}

	\begin{table}[hbt!]
	\centering
	\resizebox{1\textwidth}{!}{
		\begin{tabular}{|c c c c c|c c c c c|c c c c c|}
			\hline
			\multicolumn{5}{|c}{Meson} & \multicolumn{5}{|c|}{Baryon} & \multicolumn{5}{c|}{Tetraquark}\\
			$J^{P(C)}$	& Name & $M_{\parallel}$ & $M_\perp$  & $M$ & $J^{P(C)}$	& Name & $M_{\parallel}$  & $M_\perp$ & $M$ & 	$J^{P(C)}$	& Name & $M_{\parallel}$ & $M_\perp$  & $M$ \\
			\hline
			$0^{-+}$ & $\eta_{c}(2984)$ & 2927 & 0  & 2927 & - & - & - & - &-&-&-&- &-&-\\	
			$1^{+-}$ & $h_{c}(3525)$ & 3440 & 1046 & 3596 & $(1/2)^+$ & $\Xi^{\rm SELEX}_{cc}(3520)$ & 3254 & 1046 & 3418 &  $0^{++}$ & $\chi_{c0}(3415)$ & 3864  & 1046  & 4003  \\
			&	&   & &   & & $\Xi^{\rm LHCb}_{cc}(3620)$ & & &  &   & & & &\\
			\hline
			$1^{--}$ & $J/\psi(3096)$ & 2927 & 740 & 3019  & - & - & - & -  &-&-&-&-&-&-\\
			$2^{++}$ & $\chi_{c2}(3556)$ & 3440 & 1281 & 3671 &  $(3/2)^+$ & $\Xi^{\rm LHCb}_{cc}(3620)$ & 3254  & 1281  & 3497 & $1^{++}$ &$\chi_{c1}(3510)$ & 3580 & 1281 &  3802  \\
			\hline
			$1^{--}$ &	$\psi^\prime(3686)$ & 3440  & 1281 &  3671 & - & - &- & - &-&-&-&-&-&-\\
			$2^{++}$ & $\chi_{c2}(3927)$ & 3794 & 1654 & 4139 &  $(3/2)^+$ & $\Xi^{*}_{cc}?$ & 3751  & 1654  & 4099 & $1^{++}$ & $X(3872)$ & 4062 & 1654 & 4386\\	
			&  &  &  &  & & &  &  & & $1^{+-}$ &$Z_c(3900)$ & 4240  & 1654  & 4551 \\
			\hline
			\hline
			$0^{-+}$ & $\eta_{b}(9400)$ & 9424 & 0  & 9424 & - & - & - & -&-&-&-&-&-&-\\
			$1^{+-}$ & $h_{b}(9900)$ & 9776  & 1046  & 9832 &  $(1/2)^+$ & $\Xi_{bb}?$ & 9750 & 1046  & 9806  & $0^{++}$ & $\chi_{b0}(9860)$ & 10285 & 1046 & 10338\\
			\hline
			$1^{--}$&	$\Upsilon(9460)$ & 9424  & 740  & 9453 &  - & - & -  &-&-&-&-&-&-&-\\
			$2^{++}$ & $\chi_{b2}(9910)$ & 9776 & 1281  & 9860 & $(3/2)^+$ & $\Xi_{bb}?$ & 9750 & 1281  & 9834  & $1^{++}$ & $\chi_{b1}(9893)$ & 10081  & 1281  & 10162  \\
			\hline
			$1^{--}$	&$\Upsilon(2S)(10020)$ & 9776  & 1281 & 9860 & - & - & - & -&-&-&-&-&-&-\\
			$2^{++}$ & $\chi_{b2}(10270)$ & 10024 & 1654  & 10160  & $(3/2)^+$ & $\Xi_{bb}?$ & 10100 & 1654  & 10234 & $1^{++}$ & $X_{b}?$ & 10425  & 1654  & 10555\\
			\hline 
	\end{tabular}}
\caption{Computed masses (in MeV), using the 't Hooft potential, for hadrons with two heavy quarks, compared to the PDG data \cite{Zyla:2020zbs}.}
\label{Tab:heavy}
\end{table}

	\begin{table}[hbt!]
		\centering
		\resizebox{1\textwidth}{!}{
			\begin{tabular}{|c c c c c |c c c c c |c c c c c|}
				\hline
				\multicolumn{5}{|c}{Meson} & \multicolumn{5}{|c|}{Baryon} & \multicolumn{5}{c|}{Tetraquark}\\
				$J^{P(C)}$ & Name  & $M_{\parallel}$ & $M_\perp$ & $M$ & $J^{P(C)}$ & Name & $M_{\parallel}$ & $M_\perp$ & $M$ & $J^{P(C)}$ & Name & $M_{\parallel}$ & $M_\perp$ & $M$\\
				\hline
				$0^{-+}$ & $\pi(140)$ & 140 & 0 & 140 & - & - & - & - & - & - &-&- &-&-\\
				$1^{+-}$ & $h_{1}(1170)$ & 387 & 1046 & 1115 & $(1/2)^+$ & $N(940)$ &  189 & 1046 & 1063 & $0^{++}$ & $f_0(500)$ & 359 & 1046 & 1106\\
				$2^{-+}$ &$\eta_{2}(1645)$ & 629 & 1479 & 1607 & $(3/2)^-$ & $N(1520)$ & 312 & 1479 & 1512 & $1^{-+}$ & - & 480 & 1479 & 1555 \\
				\hline
				$1^{--}$	&$\rho(770)$, $\omega(780)$ & 140 & 740 & 753 & - & - & - & - & - & - &-&-&-&-\\
				$2^{++}$ &$a_{2}(1320)$, $f_{2}(1270)$ & 387 & 1281 & 1338 & $(3/2)^+$ & $\Delta(1232)$ & 189 & 1281 & 1295 & $1^{++}$ & $a_{1}(1260)$ &  236 & 1281 & 1302 \\
				$3^{--}$ & $\rho_{3}(1690)$, $\omega_{3}(1670)$ &  629 & 1654  & 1770 & $(3/2)^-$ & $\Delta(1700)$ & 312 & 1654 & 1683 & $1^{-+}$ & $\pi_{1}(1600)$ & 359 & 1654 & 1692 \\
				$4^{++}$ & $a_{4}(1970)$, $f_{4}(2050)$ & 870 & 1957 & 2142 & $(7/2)^+$ & $\Delta(1950)$ & 434 & 1957 & 2004 & & - & - & - &  -\\
				\hline
				$0^-$	& $\bar{K}(495)$ & 459 & 0 & 459 &- & - & - & - & - & -&-&-&-&-\\
				$1^+$&	$\bar{K}_{1}(1270)$ & 580 & 1046 & 1196 & $(1/2)^+$ & $\Lambda(1115)$ & 505 & 1046  & 1162 & $0^+$  &  $K^{*}_{0}(1430)$  & 552 & 1046 & 1183 \\
				$2^-$	& $K_{2}(1770)$ & 700 & 1479 & 1636 & $(3/2)^-$ & $\Lambda(1520)$ &  626 & 1479 & 1606 & $1^-$ & - &  672 & 1479 & 1624 \\
				\hline
				$0^-$	& $\bar{K}(495)$ & 459 & 0 & 459 & - & - & - & - & - & -&-&-&-&-\\
				$1^+$ &	$\bar{K}_{1}(1270)$ & 580 & 1046 & 1196 & $(1/2)^+$ & $\Sigma(1190)$ &  505 & 1046 & 1162 & $0^{++}$ & $a_{0}(980)$ & 984 & 1046 & 1436 \\
				&  &  &  & & & & & & &  & $f_{0}(980)$ & & & \\
				\hline
				$1^-$ & $K^{*}(890)$ & 459 & 740  & 871 &  - & -&  - & -&- & -& -&-&-&- \\
				$2^+$	&$K^{*}_{2}(1430)$ & 580 & 1281 & 1406 & $(3/2)^+$ & $\Sigma(1385)$ & 505 & 1281 & 1377 & $1^+$ & $K_{1}(1400)$ & 552 & 1281 & 1395 \\
				$3^-$	& $K^{*}_{3}(1780)$ & 700 & 1654 & 1796 & $(3/2)^-$ & $\Sigma(1670)$ & 626 & 1654 & 1768 & $2^-$ & $K_{2}(1820)$ & 672 & 1654 & 1785 \\
				$4^+$ &	$K^{*}_{4}(2045)$ & 821 & 1957 & 2122 & $(7/2)^+$ &$\Sigma(2030)$ & 747 & 1957 & 2095 & - & -&-&-&-\\
				\hline
				$0^{-+}$ & $\eta^\prime(958)$ & 772 & 0 & 772 &- &- &- &- & -&-&-&-&-&- \\
				$1^{+-}$ & $h_{1}(1380)$ & 1012 & 1046  & 1455 & $(1/2)^+$ & $\Xi(1320)$ & 818 & 1046 & 1328 & $0^{++}$ & $f_{0}(1370)$  & 984 & 1046 & 1436 \\
				& & & & & & & & & & & $a_{0}(1450)$ & & & \\
				$2^{-+}$ & $\eta_{2}(1870)$  & 1252 & 1479 & 1938 & $(3/2)^-$ & $\Xi(1620)$ & 938 & 1479 & 1751 & $1^{-+}$ & - & 1104 & 1479 & 1846 \\
				\hline
				$1^{--}$ & $\Phi(1020)$ & 772 & 740 & 1069 & - & - & - & - & - & - & - & - & - & -\\
				$2^{++}$	& $f'_{2}(1525)$ & 1012 & 1281 & 1632 & $(3/2)^+$ & $\Xi^{*}(1530)$ & 818 & 1281 & 1520 & $1^{++}$ & $f_{1}(1420)$  & 864 & 1281 & 1545 \\
				
				& &	&  & & & & & & & & $a_{1}(1420)$ & & & \\
				$3^{--}$&	$\Phi_{3}(1850)$  & 1252 & 1654 &  2074 & $(3/2)^-$ & $\Xi(1820)$ & 938 & 1654 & 1901 &  - & -&-&- & -\\
				\hline
				$2^{++}$ &	$f_{2}(1640)$ & 1012 & 1281 & 1632 & $(3/2)^+$ &$\Omega(1672)$ & 1129 & 1281 & 1708 & $1^+$ & $K_{1}(1650)$ &  1175 & 1281 & 1738 \\
				\hline
		\end{tabular}}
\caption{Computed masses (in MeV) using the alternative longitudinal potential given by Equation \eqref{UL-BFLQ}, for light hadrons compared to the PDG data \cite{Zyla:2020zbs}.}
\label{Tab:light-BLFQ}
\end{table}

	\begin{table}[hbt!]
	\centering
	\resizebox{1\textwidth}{!}{
		\begin{tabular}{|c c c c c|c c c c c|c c c c c|}
			\hline
			\multicolumn{5}{|c}{Meson} & \multicolumn{5}{|c|}{Baryon} & \multicolumn{5}{c|}{Tetraquark}\\
			$J^{P(C)}$	& Name & $M_{\parallel}$ & $M_\perp$ & $M$ & $J^{P(C)}$	& Name & $M_{\parallel}$ &  $M_\perp$ & $M$  &	$J^{P(C)}$	& Name &  $M_{\parallel}$ & $M_\perp$  &  $M$\\
			\hline
			$0^-$ &	$D(1870)$ & 1859 & 0 & 1859 & - & - &  - & - & - &- &-&-&-&-\\
			$1^+$ &	$D_{1}(2420)$ & 1979 &  1160 & 2294 & $(1/2)^+$ & $\Lambda_{c}(2290)$  & 2189  & 1160  & 2478 & $0^+$ & $\bar{D}^{*}_{0}(2400)$ & 2519  & 1160 & 2774 \\
			$2^-$ &	$D_{J}(2600)$ & 2099 & 1640  & 2664 & $(3/2)^-$ & $\Lambda_{c}(2625)$  & 2309 & 1640 & 2833 & $1^-$ & -  & 2639 & 1640 & 3108 \\
			\hline
			$0^-$ & $\bar{D}(1870)$ & 1859 & 0  & 1859 & - & - & - & - & - & - &- &- &- &-\\
			$1^+$ & $\bar{D}_{1}(2420)$ &  1979 & 1160 & 2294 & $(1/2)^+$ & $\Sigma_{c}(2455)$ & 2189 & 1160 & 2478 & $0^+$ & $D^{*}_{0}(2400)$ & 2519 &  1160 & 2774 \\
			\hline
			$1^-$ & $D^{*}(2010)$  & 1859  & 820 & 2032 & - & - & - & - & - & -&- &- &- &-\\
			$2^+$ & $D^{*}_{2}(2460)$ &  1979 & 1421 & 2436 & $(3/2)^+$ & $\Sigma^{*}_{c}(2520)$  & 2189 & 1421 & 2610 & $1^+$ & $D(2550)$  & 2519 & 1421  & 2892 \\
			$3^-$ & $D^{*}_{3}(2750)$ & 2099  & 1834 & 2788 & $(3/2)^-$ &$\Sigma_{c}(2800)$ & 2309 & 1834 & 2949 & - & - & - &- &-\\
			\hline
			$0^-$ & $D_{s}(1968)$ &  2029 & 0 &  2029 & - & - & - & - & - & - &- &- &- &-\\
			$1^+$ & $D_{s1}(2460)$ & 2149  & 1160  & 2442 & $(1/2)^+$ & $\Xi_{c}(2470)$ &  2359 &  1160  & 2629 & $0^+$ & $\bar{D}^{*}_{s0}(2317)$ &  2689  & 1160  & 2929 \\
			$2^-$ & $D_{s2}?$ & 2269 & 1640  & 2800 & $(3/2)^-$ & $\Xi_{c}(2815)$ & 2479  & 1640  & 2973 & $1^-$ &-& 2809   & 1640 & 3253 \\
			\hline
			$1^-$ &	$D^{*}_{s}(2110)$ &  2029 &  820 & 2189 & - & - & - & - & - & - & - &- &- &-\\
			$2^+$ & $D^{*}_{s2}(2573)$& 2149 & 1421 & 2576 & $(3/2)^+$ & $\Xi^{*}_{c}(2645)$  & 2359 & 1421  & 2754 & $1^+$ &$D_{s1}(2536)$ &  2689 &  1421 & 3042 \\
			$3^-$ & $D^{*}_{s3}(2860)$  & 2269 & 1834  & 2918 & - & - & - & - & - & - &- &- &- & -\\
			\hline
			$1^+$ & $\bar{D}_{s1}?$ &  2149 &  1160  & 2442 & $(1/2)^+$ &$\Omega_{c}(2695)$ & 2529 & 1160 & 2783 & $0^+$ & - &  2859 & 1160 & 3086 \\
			\hline
			$2^+$ & $D^{*}_{s2}?$ &  2149  & 1421  & 2576 & $(3/2)^+$ & $\Omega_{c}(2770)$ & 2529   & 1421  & 2901 & $1^+$ & -  & 2859  & 1421  & 3193 \\
			\hline \hline
			$0^-$ & $\bar{B}(5280)$ & 5060 & 0 &  5060 &-&- & - & - & - & - & - & -& -&-\\
			$1^+$ & $\bar{B}_{1}(5720)$ & 5180  & 2000  & 5552 & $(1/2)^+$ & $\Lambda_{b}(5620)$ &5390 & 2000  & 5749  & $0^+$ & $B_{J}(5732)$  & 5720 & 2000 & 6059 \\
			$2^-$ & $\bar{B}_{J}(5970)$ & 5300 & 2828  & 6007 & $(3/2)^-$ & $\Lambda_{b}(5920)$ & 5510  & 2828  & 6193 & $1^-$ &-&5840 &  2828 &  6489 \\
			\hline
			$0^-$ &$B(5280)$  & 5060 & 0 & 5060 & - & - & - & - & - & - &- &- &- &-  \\
			$1^+$ & $B_{1}(5720)$ &  5180 & 2000  & 5552 & $(1/2)^+$ & $\Sigma_{b}(5815)$  & 5390  & 2000  & 5749 & $0^+$ &$\bar{B}_{J}(5732)$ & 5720 & 2000 & 6059 \\
			\hline
			$1^-$ & $B^{*}(5325)$  & 5060 & 1414  & 5254 & - & - & - & - & - & - & - &- &- &-\\
			$2^+$	& $B^{*}_{2}(5747)$  & 5180  & 2449 & 5730 & $(3/2)^+$ &$\Sigma^{*}_{b}(5835)$  & 5390  & 2449 &  5920 & $1^+$ &$B_{J}(5840)$ & 5720 &  2449  & 6222 \\
			\hline
			$0^-$ & $B_{s}(5365)$ &  5230 & 0 & 5230 & - & - & - & - & - & -&- &- &- &-\\
			$1^+$	& $B_{s1}(5830)$ & 5350 & 2000  & 5711 & $(1/2)^+$ & $\Xi_{b}(5790)$ & 5597  & 2000 & 5908 & $0^+$ &$\bar{B}^{*}_{s0}?$ & 5890  & 2000  & 6220 \\
			\hline
			$1^-$	& $B^{*}_{s}(5415)$ & 5230 & 1414 & 5418 & - & - & - & - & - & -&- &-&-&-\\
			$2^+$ &	$B^{*}_{s2}(5840)$ &  5350 & 2449 & 5884 & $(3/2)^+$ & $\Xi^{*}_{b}(5950)$& 5560  & 2449 & 6075 & $1^+$ & $B_{s1}?$ &  5890 & 2449  & 6379\\
			\hline
			$1^+$& $B_{s1}?$ & 5350 & 2000&  5711 & $(1/2)^+$ & $\Omega_{b}(6045)$ & 5730 &  2000 & 6069 & $0^+$ &-& 6060 &   2000  & 6381\\
			\hline
	\end{tabular}}
\caption{Computed masses  (in MeV), using the alternative longitudinal potential given by Equation \eqref{UL-BFLQ}, for hadrons with one heavy quark, compared to the PDG data \cite{Zyla:2020zbs}.} 
\label{Tab:heavy-light-BLFQ}
\end{table}

	\begin{table}[hbt!]
	\centering
	\resizebox{1\textwidth}{!}{
		\begin{tabular}{|c c c c c|c c c c c|c c c c c|}
			\hline
			\multicolumn{5}{|c}{Meson} & \multicolumn{5}{|c|}{Baryon} & \multicolumn{5}{c|}{Tetraquark}\\
			$J^{P(C)}$	& Name  & $M_{\parallel}$ &  $M_\perp$  & $M$ & $J^{P(C)}$	& Name & $M_{\parallel}$  & $M_\perp$  & $M$  &	$J^{P(C)}$	& Name &  $M_{\parallel}$ & $M_\perp$  & $M$\\
			\hline
			$0^{-+}$ & $\eta_{c}(2984)$ & 2999 & 0 & 2999 & - & - & - & - & - & -&-&-&-&-\\	
			$1^{+-}$ & $h_{c}(3525)$ & 3239 & 1200 & 3454 & $(1/2)^+$ & $\Xi^{\rm SELEX}_{cc}(3520)$  & 3329  & 1200 &  3539 & $0^{++}$ & $\chi_{c0}(3415)$  & 3780 &  1200 &  3965 \\
			&	&  &  & & & $\Xi^{\rm LHCb}_{cc}(3620)$ & & &  &  &  & & &\\
			\hline
			$1^{--}$ & $J/\psi(3096)$ & 2999 & 848 & 3117 & - & - & - & - & - & - &- &- &- &-\\
			$2^{++}$ & $\chi_{c2}(3556)$ &  3239 & 1470  & 3557 & $(3/2)^+$ & $\Xi^{\rm LHCb}_{cc}(3620)$  & 3329  & 1470 & 3639 &  $1^{++}$ &$\chi_{c1}(3510)$  & 3660 & 1470 & 3944 \\
			\hline
			$1^{--}$ &	$\psi^\prime(3686)$ & 3239 & 1470  & 3557 & - & - & - & - & - & -& -& -&- & -\\
			$2^{++}$ & $\chi_{c2}(3927)$ & 3479 &  1897  & 3963 & $(3/2)^+$ & $\Xi^{*}_{cc}?$ & 3570 & 1897 & 4042 & $1^{++}$ & $X(3872)$ & 3900  & 1897 & 4337\\	
			&  &  &  &  & && & &  & $1^{+-}$ &$Z_c(3900)$ &  4020 &  1897  & 4445\\
			\hline
			\hline
			$0^{-+}$ & $\eta_{b}(9400)$ & 9400  & 0 & 9400 & - & - & - & - & - & - &- &- &- & -\\
			$1^{+-}$ & $h_{b}(9900)$ & 9640  & 2140 & 9874 & $(1/2)^+$ & $\Xi_{bb}?$ & 9730 &  2140 &  9962 & $0^{++}$ & $\chi_{b0}(9860)$ &  10180 & 2140 & 10402\\
			\hline
			$1^{--}$&	$\Upsilon(9460)$ &  9400  & 1513 & 9521 & - & - & - & - & - & -&- &- &- &-\\
			$2^{++}$ & $\chi_{b2}(9910)$ & 9640 & 2621 & 9990 & $(3/2)^+$ & $\Xi_{bb}?$ &  9730 & 2621 & 10077 & $1^{++}$ & $\chi_{b1}(9893)$ &  10060 & 2621 & 10396 \\
			\hline
			$1^{--}$	&$\Upsilon(2S)(10020)$ & 9640 & 1513 & 9990 & - & - & - & - & - & - &- &- &- &-\\
			$2^{++}$ & $\chi_{b2}(10270)$ &  9880 &  3384 & 10443 & $(3/2)^+$ & $\Xi_{bb}?$ &  9970 & 3384 &  10528 & $1^{++}$ & $X_{b}?$ &  10300 &  3384 & 10841\\
			\hline 
	\end{tabular}}
\caption{Computed masses (in MeV), using the alternative longitudinal potential given by Equation \eqref{UL-BFLQ}, for hadrons with two heavy quarks, compared to the PDG data \cite{Zyla:2020zbs}.}
\label{Tab:heavy-BLFQ}
\end{table}


 \end{document}